\documentclass[aps,prb, article,amsmath,amssymb,twocolumn,floatfix]{revtex4}
\usepackage{epsfig, color}
\usepackage{amsfonts, relsize}
\usepackage{bbold}
\usepackage{graphics}
\usepackage{subfigure}
\usepackage{graphicx}
\usepackage{hyperref}

\begin{document}

\title{Superuniversality of topological quantum phase transition and global phase diagram of dirty topological systems in three dimensions}
\author{Pallab Goswami}
\affiliation{Condensed Matter Theory Center and Joint Quantum Institute, Department of Physics, University of Maryland, College Park, Maryland 20742- 4111, USA}
\author{Sudip Chakravarty}
\affiliation{Department of Physics and Astronomy, University of
California Los Angeles, Los Angeles, CA 90095-1547, USA}

\date{\today}

\begin{abstract}
The quantum phase transition between two clean, non interacting topologically distinct gapped states in three dimensions is governed by a massless Dirac fermion fixed point, irrespective of the underlying symmetry class, and this constitutes a remarkably simple example of superuniversality. For a sufficiently weak disorder strength, we show that the massless Dirac fixed point is at the heart of the robustness of superuniversality. We establish this by considering both perturbative and nonperturbative effects of disorder. The superuniversality breaks down at a critical strength of disorder, beyond which the topologically distinct localized phases become separated by a delocalized diffusive phase. In the global phase diagram, the disorder controlled fixed point where superuniversality is lost, serves as a multicritical point, where the delocalized diffusive and two topologically distinct localized phases meet and the nature of the localization-delocalization transition depends on the underlying symmetry class. Based on these features we construct the global phase diagrams of noninteracting, dirty topological systems in three dimensions. We also establish a similar structure of the phase diagram and the superuniversality for weak disorder in higher spatial dimensions. By noting that $1/r^2$ power-law correlated disorder acts as a marginal perturbation for massless Dirac fermion in any spatial dimension $d$, we have established a general renormalization group framework for addressing disorder driven critical phenomena for fixed spatial dimension $d > 2$.

\end{abstract}


\maketitle
\section{Introduction} For  models of continuous quantum phase transitions (QPT) involving order parameter,  critical exponents are entirely determined by the symmetry of the order parameter  and the dimensionality. An interesting question is whether the critical exponents for a class of continuous quantum phase transitions can even become independent of the underlying symmetry properties. Since the corresponding fixed point gives rise to universal critical phenomena irrespective of the underlying symmetry class, the critical exponents can only depend on the dimensionality. This phenomenon will be termed as \emph{superuniversality} and as discussed below it plays an important role in the theory of topological QPTs.

The notion of \emph{superuniversality} was proposed by several authors~\cite{KLZ,Lutken,KF,Pruisken} for quantum Hall plateau transitions.  All integer quantum Hall plateau transitions of noninteracting fermions exhibit the same correlation length exponent $\nu \sim 7/3$ and dynamical scaling exponent $z=2$, which is an example of universality for the QPTs of two dimensional systems belonging to the unitary Wigner-Dyson symmetry class. In the presence of  interactions, the symmetry classification of noninteracting systems can break down, and there is no obvious reason why the integer and fractional quantum Hall plateau transitions should display the same critical properties. Motivated by the experimental observations and based on the topological Chern-Simons theories as well as the theta vacuum structure of unitary nonlinear sigma models, it has been argued that all  quantum Hall plateau transitions (irrespective of  integer or fractional filling fractions) display the same correlation length exponent $\nu \sim 7/3$ and the dynamic scaling exponent $z =1$~\cite{KLZ,Lutken,KF,Pruisken}, which is an intriguing example of \emph{superuniversality}. It has been conjectured that as a consequence, the critical conductances at different plateau transitions become  related by modular transformations~\cite{Lutken, KF}. Thus, \emph{superuniversality} is an essential ingredient of the comprehensive theory of the global phase diagram~\cite{KLZ} for  quantum Hall systems~\cite{KF,Pruisken}.

In recent years there has been a great surge of interest in three dimensional topological states of matter~\cite{MooreBalents,Roy,FuKane1,FuKane2,QiZhang1,RyuLudwig,HasanKane,QiZhang2}. In this context, a fundamental question arises if there is any \emph{superuniversality} for three dimensional topological QPTs, and if it can play any role in determining the global phase diagrams of three dimensional topological states of matter. Addressing superuniversality and constructing  global phase diagrams for dirty topological systems in three dimensions are the motivations of our present work. 

\begin{figure}[htb]
\begin{center}
\includegraphics[scale=0.65]{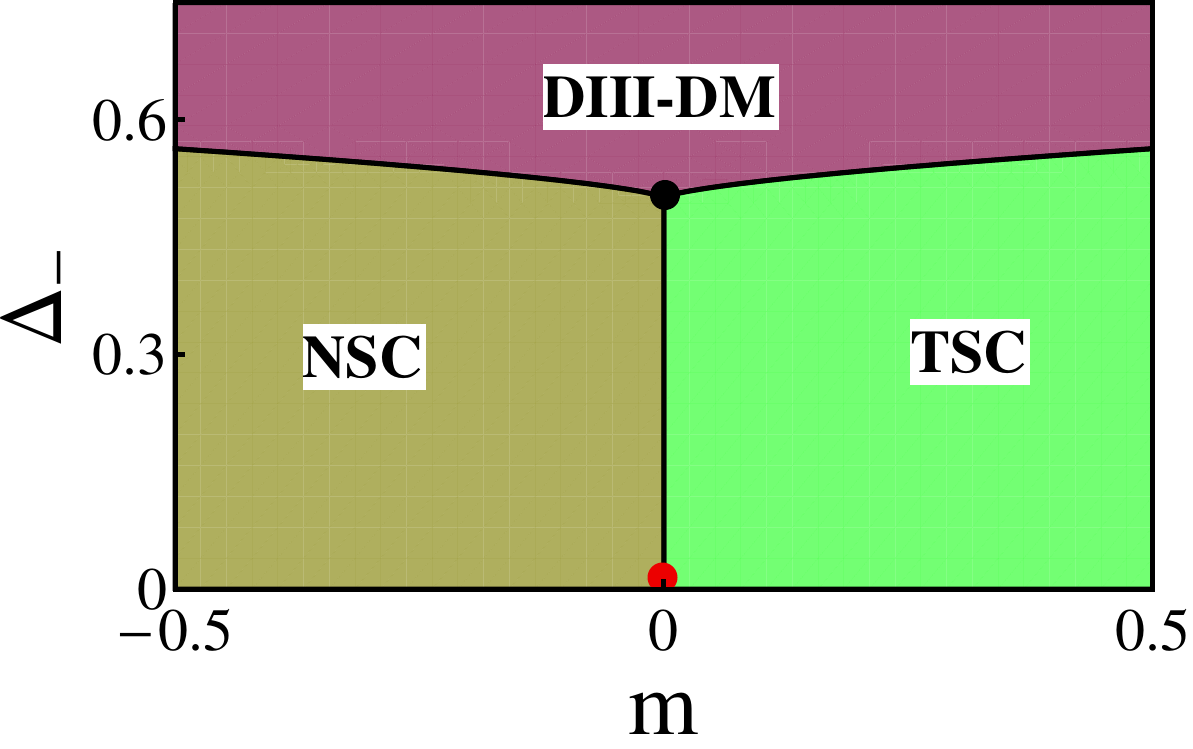}
\caption{The phase diagrams of dirty superconductors in class DIII for three dimensions. The coupling constant $m$ is the disorder averaged Dirac mass and it corresponds to the chemical potential of the quasiparticles in normal state, and $\Delta_-=\Delta_{45}-\Delta_4$ is a combination of two underlying disorder couplings, $\Delta_{4}$ and $\Delta_{45}$. For the typical example of $^3$He-B, $\Delta_{45}$ and $\Delta_4$ respectively correspond to the strengths of random, real s-wave pairing and a random chemical potential for normal quasiparticles, while for a different physical system they can represent realization of different random potentials. While the disorder couplings $\Delta_{45}$ and $\Delta_4$ are positive definite quantities, $\Delta_-$ can acquire both positive and negative values. When $\Delta_-$ is less than the critical strength $\Delta^\ast_-$, the direct transition between two topologically distinct localized states [namely the topological superconductor (TSC) and the normal superconductor (NSC)] is described by the massless, Dirac fermion fixed point (red dot) with a dynamic scaling exponent $z=1$ and a correlation/localization length exponent $\nu_M=1$ (scaling dimension of the Dirac mass $m$, describing a correlation/localization length $\xi_M \sim 1/|m|$). The disorder controlled multicritical point is marked by the black dot, where the TSC, NSC and the diffusive phase DIII-DM meet. We note that the DIII-DM also describes a paired state, and not an ordinary diffusive metal or normal state. At the multicritical point, the correlation/localization length $\xi_M$ diverges with an exponent $\nu_M=2/3$ for Gaussian white noise disorder, while the correlation length exponent for disorder (the mean free path of the diffusive phase) $\xi_\Delta$ diverges with an exponent $\nu_\Delta$. For Gaussian white noise disorder and at one loop level we find $\nu_\Delta = 1$. The multicritical point is actually a projection of a line of fixed points on the $m - \Delta_-$ plane, determined by $\Delta_-=\Delta^\ast_-$ and the arbitrary value of $\Delta_+=\Delta_{45}+\Delta_4$. Along the line of fixed points, the correlation length exponents have universal values, while the dynamical exponent varies continuously depending on $\Delta_{+}$ (or the ratio $\Delta_{45}/\Delta_4$). The other four symmetry classes AIII, AII, CI, CII also possess similar phase diagrams as function of disorder strength and Dirac mass. Dirty topological systems in higher spatial dimensions also exhibit similar structure of the phase diagram, irrespective of the underlying symmetry class.}
\label{phasediagram}
\end{center}
\end{figure}

The Altland Zirnbauer (AZ)~\cite{AltlandZirnbauer,EversMirlin} classification (Table~\ref{table1}) of random matrices plays an important role in obtaining the topological distinction for localized states of noninteracting fermions~\cite{RyuLudwig}. In the absence of disorder, a four component massless Dirac fermion describes the transition between two topologically distinct gapped states~\cite{Volovik,Liu,Murakami,GoswamiChakravarty,NomuraRyu} in the symmetry classes AII, AIII and DIII. By contrast, an eight component massless Dirac fermion is required for a minimal model of critical excitations~\cite{RyuLudwig,NomuraRyu} for describing the QPTs in the symmetry classes CI and CII. The crux of our idea is that the massless Dirac Hamiltonian describes a quantum critical system with a dynamic scaling exponent $z=1$ and a correlation length exponent $\nu_M=1$ (scaling dimension of Dirac mass $m$ which describes a length scale $\xi_M=\hbar v/|m|$, with $v$ being the Fermi velocity), and from this simple observation~\cite{GoswamiChakravarty} several critical properties can be obtained for all five AZ symmetry classes. For sufficiently weak disorder in $d=3$, we show within a one-loop renormalization group approximation that the \emph{superuniversality} involving massless Dirac fixed point remains unaffected, but it eventually breaks down at a disorder controlled multicritical point, where two insulating/localized states and the diffusive metal meet, as shown in Fig.~\ref{phasediagram}.

\begin{table}[htbp]
\begin{center}
\begin{tabular}{|l|l|l|l|l|l|}
\hline
Class & TRS & SRS  & d=1 & d=2 & d=3\\
\hline
\hline
GUE/ A & - & $\pm$ & - & $Z$ & - \\ \hline
GOE/ AI & + & + & - & - & - \\ \hline
GSE/ AII & + & - & - & $Z_2$ & $Z_2$ \\ \hline
CUE/ AIII & - & $\pm$ & $Z$ & - & $Z$ \\ \hline
COE/ BDI & + & + & $Z$ & - & - \\ \hline
CSE/ CII & + & - & $2Z$ & - & $Z_2$ \\ \hline
BdG, D & - & - & $Z_2$ & $Z$ & -  \\ \hline
BdG, C & - & + & - & $2Z$ & - \\ \hline
BDG, DIII & + & - & $Z_2$ & $Z_2$ & $Z$ \\ \hline
BdG, CI & + & + & - & - & $2Z$ \\ \hline \hline
\end{tabular}
\end{center}
\label{table1}
\caption{The Altland-Zirnubauer symmetry classes and allowed topological invariants for distinguishing gapped phases within a given class. By GUE, GOE, GSE, CUE, COE, CSE we respectively denote Gaussian unitary, Gaussian orthogonal, Gaussian symplectic, chiral unitary, chiral orthogonal and chiral symplectic ensembles. Here TRS and SRS respectively denote time reversal symmetry and SU(2) spin rotational symmetry. In the second and third columns the presence (absence) of a certain symmetry is denoted by $+$ ($-$) symbol. The absence of a topological invariant is also denoted by $-$ symbol in the fourth, fifth and sixth columns. For two classes A and AIII, the $\pm$ symbol in the third column signifies that the SRS symmetry may or may not be present for these two AZ classes. First three classes follow from the conventional `Wigner-Dyson' classification. The next three chiral classes are obtained by imposing an additional discrete sublattice or chiral symmetry on the `Wigner-Dyson" classes. The last four classes describe the Bogoliubov-de Gennes (BdG) quasiparticles of superconducting or superfluid systems that break particle-number conservation. Under special circumstances two unitary classes A and AIII can also describe superconducting systems. For example class AIII can describe a time-reversal invariant superconducting state which breaks SU(2) SRS but preserves a U(1) subgroup of it~\cite{RyuLudwig}.}
\end{table}

There is some experimental evidence for quantum critical points separating spin-orbit coupled topological and trivial insulators belonging to class AII in Bi$_{1-x}$Sb$_x$~(Ref.~\onlinecite{Lenoir,Goswami2,Teo}), BiTl(S$_{1-\delta}$Se$_\delta$)$_2$~(Ref.~\onlinecite{Hasan1,Sato1}) and (Bi$_{1-x}$In$_x$)$_2$Se$_3$~(Ref.~\onlinecite{Brahlek,Wu,Hasan2}). The quantum critical Dirac semimetal (with odd number of Dirac cones) is distinct from the symmetry (inversion, time reversal and discrete rotation) protected topological Dirac semimetal phase with two Dirac cones as observed in Cd$_3$As$_2$ and Na$_3$Bi~(Ref.~\onlinecite{Wang, Neupane,Borisenko,LiuChen1,LiuChen2,Xu}).

The topological states of matter, as modeled by massive Dirac fermions, are also of great interest in high energy physics~\cite{Witten1,Witten2}. The momentum dependent Dirac mass term in effective low energy models of condensed matter physics is nothing but the Wilson mass used in lattice gauge theory for avoiding fermions doublers~\cite{Nielsen}. The existence of protected chiral fermions on the surface of a topological insulator in (4+1) dimensions was originally conceived in Ref.~\onlinecite{Kaplan1} as a way of simulating chiral fermions in lattice gauge theory. More recently in Ref.~\onlinecite{Kaplan2}, space-time has been proposed as a topological insulator in (4+1) dimensions for explaining the mechanism behind the existence of three particle generations in the standard model. In addition, the notion of bulk-boundary correspondence and topological response functions (such as quantized magnetoelectric effect in AII topological insulator) have intimate connections with the axial anomaly of relativistic fermions~\cite{Jackiw,Callan,Fujikawa,ZhangGrav,RyuMoore}. Even though massive and massless Dirac fermions have been considered as the building blocks of particle physics, all the fermions in high energy physics are actually massive in the low energy limit, due to the chiral symmetry breaking caused by strong interactions. Therefore, in the context of particle physics, the massless Dirac fermions are only relevant at high energy or temperature scales. By contrast, the massless Dirac fermions in condensed matter physics are realized as emergent excitations in the low energy limit. Therefore, it is becoming feasible to study the exotic low energy physics of massless Dirac fermions by using table top experiments, and many ideas of high energy physics have also become germane to condensed matter physics.

We note that for any underlying symmetry class, the three dimensional diffusive metal describes a stable phase, as suggested by the analysis of appropriate nonlinear sigma model of diffusive modes for each AZ class~\cite{EversMirlin}. The word diffusive phase/metal for a general AZ class should be understood with some caution. The disorder induced diffusive phase for a BdG class (D, DIII, C, CI) shows constant density of states at zero energy, $T$ linear specific heat and thermal conductivity just like a conventional diffusive Fermi liquid. However, a diffusive BdG phase still describes a paired state as opposed to a regular metal, since the global U(1) symmetry remains broken as a defining criterion for any BdG symmetry class. For symmetry classes DIII, AIII, CI and CII supporting only particle-hole symmetric Gaussian white noise disorder, we predict the localization length exponent at the multicritical point within one-loop approximation to be $\nu_M=2/3$, which saturates the bound provided by the Chayes-Chayes-Fisher-Spencer (CCFS) theorem~\cite{CCFS}. For the class AII (lacking particle hole symmetry), we show a similar relation for the dynamic scaling exponent $z$ being $3/2$ at the multicritical point.

Previously we have addressed the effects of generic time-reversal invariant disorder and Coulomb interactions on the topological QPT in three spatial dimensions for class AII~\cite{GoswamiChakravarty}. Beyond the critical disorder strength associated with a semimetal to metal QPT~\cite{Murakami,GoswamiChakravarty,NomuraRyu,Fradkin}, the topologically distinct insulating states become separated by a diffusive metallic phase, allowing only metal-insulator transitions. The perturbative calculations of the critical properties for Gaussian white noise disorder within the one loop approximation (leading order in $d=2+\epsilon$ expansion) {\em suggest}
\begin{equation}
z=3/2, \: \: \nu_\Delta=1, \: \: \nu_M=2,
\label{eq1}
\end{equation}
where $\nu_\Delta$ is the correlation length exponent of the disorder (describes the divergence of the mean free path inside the diffusive metal). The predictions regarding the structure of the phase diagram and the dynamic scaling exponent have been approximately confirmed through nonperturbative numerical analysis\cite{Kobayashi1,Kobayashi2}. Recently, the related semimetal-metal transition for Dirac and Weyl semimetals are being intensively studied through analytical~\cite{RoyDasSarma,Syzranov1,Syzranov2,Syzranov3,Nandkishore} and numerical calculations~\cite{Pixley1,LiuShindou,Sbierski1,Sbierski2,Pixley2,Bera,Hughes,Pixley3}. Even though there is a general agreement regarding $z \sim 3/2$ (within a few percent accuracy), there are uncertainties over the precise value of $\nu_\Delta$. However to the best of our knowledge, the numerical determination of $\nu_M$ is still an open problem.

Our arguments and calculations in $d=3$ can be immediately generalized to the dirty topological states or Dirac fermions in any dimension $d > 3$. Due to the stability of the massless Dirac fixed point for any $d \geq 3$ (for white noise distribution), the phase diagrams of dirty topological states for all symmetry classes in any dimension $d \geq 3$ should be similar. Some interesting higher dimensional examples are four dimensional quantum Hall plateau transition (class A), and the transition between four dimensional class AII topological and trivial insulators ($Z$ invariant). We have performed an explicit one loop RG calculation for class AII model in $d=4$ to justify our claims and found that the dynamic scaling exponent $z=d/2=2$ at the multicritical point for Gaussian white noise disorder. We also find that the disorder correlation length exponent for Gaussian white noise disorder becomes $\nu_\Delta=1/2$ (satisfying CCFS bound), indicating a mean-field nature of the transition. {\em This shows $d=4$ to be the upper critical dimension for such transitions, in contrast to the conventional Anderson localization transition, which does not possess an upper critical dimension.} 

Since topological aspects are sensitive to the underlying spatial dimensionality, it is crucial to develop a renormalization group method that does not rely on dimensional continuation scheme. In this paper we develop such renormalization group calculations for addressing disordered systems belonging to different AZ symmetry classes in $d \geq 3$, and make new predictions for disorder driven critical phenomena. For any spatial dimension $d$, the disorder distribution with $1/r^2$ power law correlation acts as a marginal perturbation for massless Dirac fermions. Therefore, the renormalization group analysis of disorder effects can be controlled by working with respect to this marginal probability distribution. We will introduce a general $1/r^{(d-\alpha)}$ power-law correlated disorder for $d>2$. The marginal $1/r^2$ distribution is obtained for $\alpha_m=(d-2)$, and the Gaussian white noise distribution is recovered by setting $\alpha=0$. (I) By holding $d$ fixed, we can define an expansion scheme $\alpha=\alpha_m-\epsilon$ to control the perturbative analysis. Thus, we will be working with a $1/r^{2+\epsilon}$ power-law correlated disorder at a fixed spatial dimension.  Within this scheme, all loop integrals would be performed for $\alpha=\alpha_m$ for extracting the logarithmic divergence and at the end of the calculations we would set $\epsilon=\alpha_m-\alpha$. Notice that by taking the formal limit, $\epsilon \to \alpha_m=(d-2)$ (or $\alpha=0$) one can obtain the results for white noise distribution. (II) Alternatively, for a fixed $\alpha$, we can define a marginal spatial dimension $d_m=\alpha+2$, and use an expansion scheme $d=d_m+\epsilon$. This is a generalization of of conventional $d=2+\epsilon$ expansion for Gaussian white noise disorder. Notice that we are then considering $1/r^{2+\epsilon}$ power-law correlations by keeping a fixed $\alpha$. All loop integrals are then carried out at $d=d_m$ and at the end of the calculations one has to set $\epsilon=(d-2)$. A priori it is not clear, which method at the leading order will provide a better description of the disorder driven critical phenomena. Here, we will employ $\alpha=\alpha_m-\epsilon$ expansion scheme for describing the disorder effects on topological quantum phase transitions for fixed spatial dimensions.

In the context of scalar or axial chemical potential disorder driven semimetal-metal transition at $d=3$, our one loop analysis within the scheme (I) leads to
$$z=1+\epsilon/2, \; \nu_\Delta=1/\epsilon, \; \eta_\psi=3\epsilon/8,$$ where $\epsilon=1-\alpha$, $\eta_\psi$ is the anomalous scaling dimension of the fermion field. Consequently, upon approaching the critical point from the semimetal side, the quasiparticle residue of massless Dirac fermion vanishes according to $Z \sim \delta^{\nu_\Delta \eta_\psi}=\delta^{3/8}$ (with $\delta$ being the reduced distance from the critical point) irrespective of $\alpha$, when $\alpha<1$. We also find that the zero-frequency, critical fermion propagator $G(0,\mathbf{k}) \sim |\mathbf{k}|^{(\eta_\psi-z)} = |\mathbf{k}|^{-(1+\epsilon/8)}$. After taking the formal limit $\epsilon \to 1$ we find 
$$z=3/2, \; \nu=1, \; \eta_\psi=3/8,$$ such that $G(0,\mathbf{k}) \sim |\mathbf{k}|^{-9/8}$, for Gaussian white noise distribution. These can be contrasted with the following predictions of one loop analysis within the $d=2+\epsilon$ expansion scheme $$z=3/2, \; \nu=1, \eta_\psi=(z-1)=1/2,$$ such that the quasiparticle residue vanishes as $Z \sim \delta^{1/2}$, but the critical propagator behaves as $G(0,\mathbf{k}) \sim |\mathbf{k}|^{-1}$ akin to the clean model. At the scalar potential disorder controlled multicritical point for class AII topological insulators, this method predicts $\nu^{-1}_{M}=1-\epsilon/4$. Therefore, for Gaussian white noise distribution we find $\nu_M=4/3$ in contrast to $\nu_M=2$ obtained from $d=2+\epsilon$ expansion~\cite{GoswamiChakravarty}.

Our paper is organized as follows: in Sec.~\ref{sec:masslessDirac} we consider the clean, noninteracting Hamiltonian of different symmetry classes and elucidate the notion of \emph{superuniversality} of massless Dirac fermion fixed point. In Sec.~\ref{sec:Harris} we formulate a generalized Harris criterion for the stability of massless Dirac fixed point against different types of disorder. This formulation can be applicable for assessing the stability of both repulsive and attractive fixed points. The subgap states induced by weak disorder and their localized nature are discussed in Sec.~\ref{sec:rare}. In Sec.~\ref{sec:LD} we present perturbative one loop calculations for obtaining a qualitative description of the multicritical point and localization-delocalization transition. The issue of superuniversality and its breakdown at the disorder controlled multicritical point in $d=4$ are addressed in Sec.~\ref{sec:superd4}. The scaling properties of physical quantities for three dimensional systems belonging to the symmetry classes DIII, AIII and AII are discussed in Sec.~\ref{sec:scaling}. We summarize our findings in Sec.~\ref{sec:conclusions}. The technical details on the symmetry properties of random axial chemical potential and the derivation of the RG flow equations are respectively provided in Appendix~\ref{sec:AppendixA} and Appendix~\ref{sec:AppendixB}. Some simple methods of obtaining the qualitative forms of the scaling functions for class DIII are described in Appendix~\ref{sec:AppendixC}.

\section{Topology and Massless Dirac fixed point}\label{sec:masslessDirac}
In this section we consider toy models of three dimensional topological states in five different symmetry classes in terms of low energy massive Dirac fermions. This will set the ground for clarifying the superuniversality of topological transitions in clean, gapped systems. We will explicitly consider the models pertinent to experimentally realized topological systems such as superfluid $^3$He-B phase (class DIII), topological insulator Bi$_2$Se$_3$ (class AII) and also class AIII (yet to be experimentally realized), which can be understood in terms of a four component Dirac fermion. 

\subsection{Class DIII superfluid} The reduced BCS Hamiltonian for the B phase is given by
\begin{eqnarray}\label{3dH0}
H_{1}=\frac{1}{2}\int \frac{d^3k}{(2\pi)^3} \; \Psi^\dagger_{\mathbf{k}} \hat{h}_{1}(\mathbf{k}) \Psi_{\mathbf{k}},
\end{eqnarray}
where $\Psi^\dagger_{\mathbf{k}}=(c^\ast_{\mathbf{k},\uparrow}, c^\ast_{\mathbf{k},\downarrow},c_{-\mathbf{k},\uparrow},c_{-\mathbf{k},\downarrow})$ is the four component Nambu spinor, $c_{\mathbf{k},s}$ is the annihilation operator for a normal state quasiparticle or fermionic $^3$He atom with spin projection $s=\uparrow / \downarrow$. The operator $\hat{h}_{1}(\mathbf{k})=\sum_{j=1}^{4} \; n_j(\mathbf{k}) \Gamma_j$, where we have introduced a four component vector
\begin{equation}
\mathbf{n}(\mathbf{k})=(\hbar vk_x,\hbar vk_y, \hbar vk_z,-\mu + \hbar^2k^2/(2m^\ast)),
\end{equation} with $\mu$ and $m^\ast$ respectively being the chemical potential and the effective mass of the normal quasiparticles. The velocity $v=\Delta_t/(\hbar k_F)=\Delta_t/\sqrt{2 m^\ast |\mu|}$ with $\Delta_t$ being  the triplet pairing amplitude, and $\Gamma_j$s are four mutually anticommuting Dirac matrices: $\Gamma_1=-\sigma_3 \otimes \tau_1$, $\Gamma_2=-\sigma_0 \otimes \tau_2$, $\Gamma_3=\sigma_1 \otimes \tau_1$, $\Gamma_4=\sigma_0 \otimes \tau_3$. The Pauli matrices $\sigma_\mu$ and $\tau_\mu$ respectively operate on the spin and particle-hole  indices. Therefore, the quasiparticles of the gapped B phase are four component massive Dirac fermions with a momentum dependent mass $m_{\mathbf{k}}=-\mu + \hbar^2k^2/(2m^\ast)$. As a defining criterion for class DIII, the Hamiltonian possesses a discrete particle-hole symmetry $\{\hat{h}_{1}, \Gamma_5 \}=0$, where $\Gamma_5=\sigma_2 \otimes \tau_1$. If $\psi$ is an eigenstate of $\hat{h}_{0}(\mathbf{k})$ with energy $E$, $\Gamma_5 \psi$ is an eigenstate with energy $-E$. Due to the Pauli principle we cannot multiply $\Psi^\dagger \Gamma_5 \Psi$ by any odd power of momentum. If we added the term $\Psi^\dagger \Gamma_5 \Psi$, the resulting $p+is$ paired state will belong to class D, as it simultaneously breaks time reversal and parity symmetries~\cite{GoswamiRoy}. The $p+is$ paired state describes a dynamical axionic state of Majorana fermions.

Inside the BCS or B phase with $\mathrm{sgn}(\mu m^\ast)>0$, the SO(4) unit vector $\hat{n}(\mathbf{k})=\mathbf{n}(\mathbf{k})/|\mathbf{n}(\mathbf{k})|$ has the form of an instanton in the three Euclidean space, as $\hat{n}(k=0)=(0,0,0,-1)$ and $\hat{n}(k=\infty)=(0,0,0,1)$. The topological invariant of the instanton is determined by the homotopy $\Pi_3(S^3)=Z$, and consequently the B phase is possesses the nontrivial topological invariant given by $N= \mathrm{sgn} (v)$. By contrast, inside the topologically trivial BEC phase with $\mathrm{sgn}(\mu m^\ast)<0$, the instanton profile is absent, as the unit vector points along $(0,0,0,1)$ both at $k=0$ and $k=\infty$. Consequently, the topological distinction between the BEC and BCS states arises through the sign of the uniform Dirac mass or the chemical potential of the normal quasiparticles, and a topological quantum phase transition occurs between these two phases at $\mu=0$. Due to the presence of momentum dependent mass term $\hbar^2k^2/(2m^\ast) \Gamma_4$, the Hamiltonian at the quantum critical point still possesses the discrete particle-hole symmetry of class DIII.

\subsection{Class AIII}
Following Ref.~\onlinecite{RyuLudwig}, the low energy block off-diagonal Hamiltonian operator for class AIII can be written as
\begin{equation}
\hat{h}_2=\hbar v \sum_{j=1}^{3} \; \Gamma_j k_j + \Gamma_5 (M_0 -B k^2),
\end{equation}
where we are using the conventional Dirac matrices $\Gamma_j=\sigma_j \otimes \tau_1$ for $j=1,2,3$, $\Gamma_4=\sigma_0 \otimes \tau_3$ and $\Gamma_5= \sigma_0 \otimes \tau_2$. The condition of chirality or discrete particle-hole symmetry is determined by $\{ \hat{h}_2, \Gamma_4 \}=0$. Due to this condition we cannot allow any block-diagonal term that can be written with identity, $\Gamma_4$, $i\Gamma_4\Gamma_5\Gamma_j$ and $i\Gamma_5\Gamma_j$ (with j=1,2,3) matrices in the Hamiltonian. Similar to the class DIII, the integer topological invariant for class AIII is determined by the homotopy $\Pi_3(S^3)=Z$ and the topological transition occurs at $M_0=0$. Notice that the class DIII Hamiltonian can also be written in the form announced for class AIII after performing a suitable unitary transformation. An important difference between these two classes is the precise nature of the quasiparticles. While class DIII involves only real or Majorana fermions, class AIII involves complex fermions.

\subsection{Class AII topological insulator}
Up to the cubic order, the low energy $\mathbf{k} \cdot \mathbf{p}$ Hamiltonian around the $\Gamma$ point for $Z_2$ topological insulator Bi$_2$Se$_3$ is given by
\begin{eqnarray}
H_3=\int \frac{d^3k}{(2\pi)^3} \; \chi^\dagger_{\mathbf{k}}\hat{h}_{3}(\mathbf{k}) \Psi_{\mathbf{k}},
\end{eqnarray}
where $\chi^\dagger_{\mathbf{k}}=(c^\ast_{+,\uparrow}, c^\ast_{-,\uparrow}, c^\ast_{+,\downarrow}, c^\ast_{-,\downarrow})$, and $\pm $ respectively denote even and odd parity orbitals. The Hamiltonian operator
\begin{eqnarray}
\hat{h}_3=\epsilon (\mathbf{k}) \mathbb 1  + A(k_\perp)(\Gamma_1 k_y -\Gamma_2 k_x) + B(k_z) k_z \Gamma_3 
\nonumber \\ + M(\mathbf{k}) \Gamma_4 + R_1(3k^2_xk_y-k^3_y)\Gamma_3  + R_2(3k^2_yk_x-k^3_x) \Gamma_5,
\end{eqnarray}
where $\Gamma_1=\sigma_1 \otimes \tau_1$, $\Gamma_2=\sigma_2 \otimes \tau_1$, $\Gamma_3=\sigma_0 \otimes \tau_2$, $\Gamma_4=\sigma_0 \otimes \tau_3$ and $\Gamma_5=\sigma_3 \otimes \tau_1$ are five mutually anticommuting $4 \times 4$ gamma matrices, and $\mathbb 1$ is the $4\times4$ identity matrix. In contrast to the class DIII Hamiltonian, we have a $\Gamma_5$ matrix with odd powers of momentum, as well as an identity matrix. Here these terms are allowed as only the time reversal symmetry has to be respected and there is no special requirement of particle-hole symmetry for class AII. The asymmetry between conduction and valence bands is captured by $\epsilon (\mathbf{k})=C_0+C_1 k^2_z+C_2k^2_\perp$, which does not have any role in determining the topology. The other band parameters $A(\mathbf{k}_\perp)=A_0+A_1 k^2_\perp$, $B(k_z)=B_0+B_1 k^2_z$, $k^2_\perp=k^2_x+k^2_y$, $M(\mathbf{k})=M_0+M_1 k^2_z+ M_2k^2_\perp$ with $M_1M_2>0$. The topologically trivial and nontrivial states can be understood by keeping the terms up to the quadratic order, and they respectively correspond to $\mathrm{sgn}(M_0 M_j)<0$ and $\mathrm{sgn}(M_0 M_j)>0$. Consistent with the underlying rhombohedral crystalline symmetry, the cubic terms proportional to $R_1$ and $R_2$ break the O(2) rotational symmetry in $k_x-k_y$ plane down to a three fold rotational symmetry. As $R_2 (3k^2_yk_x-k^3_x) \Gamma_5$ breaks the discrete particle hole symmetry defined for class DIII, it produces a hexagonal warping in the dispersion relation of the surface states. At the topological quantum phase transition the uniform Dirac mass $M_0$ vanishes, and the $k^2$ and $k^3$ terms emphasize that the problem still belongs to the class AII. For simplicity hereon we will ignore the $\epsilon(\mathbf{k})$ term, as it does not affect the topological aspects.

\begin{figure}[htb]
\begin{center}
\includegraphics[scale=0.65]{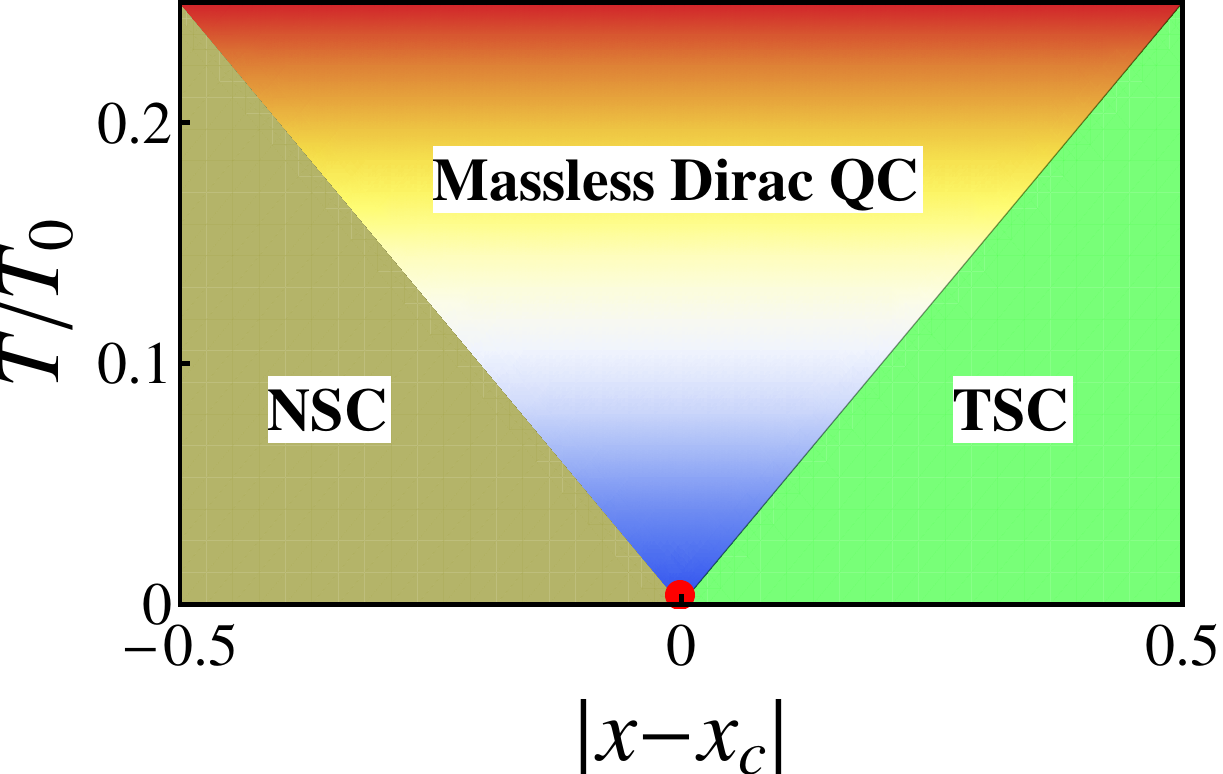}
\caption{Quantum critical fan for the direct quantum phase transition between topologically distinct localized states. In the absence of disorder, the topologically distinct insulating states possess sharp spectral gaps, described by the Dirac mass. In the presence of disorder, there is no sharp spectral gap, and the inverse of the disorder averaged Dirac mass describes the localization length ($\xi_M \sim 1/|m|$) for the sub-gap states. The tuning parameter $x$ is a combination of disorder couplings and at $x=x_c$ the disorder averaged Dirac mass vanishes, which determines the phase boundary between two localized states. In the weak disorder regime for all five Altland-Zirnbauer symmetry classes, the average Dirac mass behaves as $\bar{m}=m(0)[1-x/x_c]$, where $m(0)$ is the Dirac mass in the absence of disorder. For class DIII, $x=\Delta_-=\Delta_{45}-\Delta_4$ and $x_c \approx 6m(0) m^\ast/(\hbar^2 \Lambda^2)$ where $\Lambda$ is the momentum cut-off for the Dirac theory. The quantum critical behavior of massless Dirac fermion is observed for $k_B T >> |x-x_c|^{z \nu_M}$, with dynamic scaling exponent $z=1$, and the localization length exponent $\nu_M=1$. This picture is only applicable for sufficiently weak disorder. For strong enough disorder, the massless Dirac fermion can undergo a semimetal to metal transition, and there will be no direct transition between two topologically distinct localized states. Here $T_0=\hbar v \Lambda/k_B$ corresponds to a microscopic energy scale, where $v$ is the velocity of the massless Dirac fermions and $k_B$ is the Boltzmann constant.}
\label{fan}
\end{center}
\end{figure}

Compared to the $\mathbf{k}$-linear terms, the momentum dependent mass term for any class (as well as cubic terms in class AII) is irrelevant in the renormalization group sense. Consequently the universal properties at low momentum for DIII ($|\mathbf{k}|< 2m^\ast v/\hbar$), AIII ($|\mathbf{k}|< \hbar v/B)$ and AII are captured by a fixed point Hamiltonian of the four component massless Dirac fermion
\begin{equation}
\hat{h}_{*}=\hbar v \mathbf{k} \cdot \boldsymbol \Gamma.
\end{equation}
In the above equation we have ignored velocity anisotropies. \emph{The fixed point Hamiltonian possesses a continuous chiral symmetry, i.e., $[\hat{h}_{*},\Gamma_4 \Gamma_5]=0$. It is the consequence of the continuous chiral symmetry that the fixed point Hamiltonian can be a member of AII, DIII as well AIII. Hence, a four component massless Dirac fermion describes a simple example of superuniversality for the topological phase transition in noninteracting, clean systems.} In the similar spirit, an eight component massless Dirac fermion describes the topological transition for classes CI and CII.

At the massless Dirac fixed point the dynamic scaling exponent is given by $z=1$. Since the anomalous dimension of the Dirac mass operator is $-1$, the correlation length exponent associated with the mass operator is given by $\nu_M=1$. This correlation length controls the localization length or the penetration depth of the zero energy surface states. The density of states for a massive Dirac fermion behaves as $\rho(E) \sim |E| \Theta(E^2-\mu^2)$, reflecting the existence of a hard spectral gap for $\mu \neq 0$, and $|\mu| > |E|$. By using this density of states for calculating free energy density we can infer that the quantum critical fan (as shown in Fig.~\ref{fan}) corresponds to the region $k_BT > |\mu|$, and the singular part of the thermodynamic free energy in this region scales according to $f \sim T^{d+1} \to T^4$. Now we are in a position to investigate the effects of weak quenched disorder.

\section{Generalized Harris criterion for the stability of massless Dirac fermion}\label{sec:Harris}

Usually the stability of a clean quantum critical point against weak disorder is understood in terms of Harris criterion~\cite{Harris}. In a conventional set up of order parameter field theory, one considers the random variation of critical coupling $g_c$. The Harris criterion implies that the clean quantum critical point is stable against infinitesimally weak disorder when the correlation length exponent $\nu_{pure} >2/d$, where $d$ is the spatial dimension. For the marginal case of $\nu_{pure} =2/d$ one cannot make any heuristic statement about the stability. In the literature often the statement of CCFS theorem~\cite{CCFS} is confused with that of heuristic Harris criterion. The precise statement of CCFS theorem is that the average correlation length exponent satisfies $\bar{\nu}>2/d$ at any stable critical point in the disordered systems. Therefore, CCFS theorem not only asserts the stability of a clean critical point when $\nu_{pure}>2/d$, it also implies that the average correlation length exponent for disorder controlled critical point satisfies $\bar{\nu} >2/d$. For assessing the stability of massless Dirac fixed point in the presence of randomly varying band gap or Dirac mass (mass disorder), we can immediately take over the Harris criterion. As the correlation length exponent $\nu_M=1$ ($>2/3$), the massless Dirac fixed point will be stable in three dimensions against weak mass disorder. However, we cannot make such a statement in the presence of any other type of disorder potential which couples to different fermion bilinear without generalizing the Harris criterion.

\subsection{Symmetry allowed disorder potentials}
Before any further discussion of Harris criterion, we first consider the types of disorder allowed by the symmetry classification. We begin with class DIII, for which the Hamiltonian describing all types of quenched disorder has to anticommute with the matrix $\Gamma_5$, as a consequence of the combined operations of time reversal and particle hole symmetries for a paired state. This restricts us to the following Hamiltonian of the generic disorder
\begin{eqnarray}\label{disorderDIII}
H_{D,1}= \int d^3x \; \Psi^\dagger(x) \; [V_4(\mathbf{x}) \Gamma_4 + V_{45}(\mathbf{x})i\Gamma_4 \Gamma_5]\Psi(x).
\end{eqnarray}
For $^3$He-B these disorder potentials physically correspond to: (i) $V_4(x)$ is a random chemical potential for normal quasiparticle that acts as the random Dirac mass, (ii) a random $s$ wave pairing with real amplitude acts as the random axial potential $V_{45}(x)$. We can also add a random variation of the triplet pairing amplitude and a random spin-orbit coupling for normal state quasiparticles respectively described by $i \sum_{j=1}^{3} V_j(x)\Psi^\dagger(x)\Gamma_j \Psi(x)$ and $\sum_{j=1}^{3} V_{5j}(x)\Psi^\dagger \Gamma_5 \Gamma_j \partial_j \Psi(x)$. Since we are dealing with a Nambu spinor, Pauli principle dictates $\Psi^\dagger \Gamma_j \Psi = 0$ and $\Psi^\dagger \Gamma_5 \Gamma_j \Psi = 0$. For this reason disorder potentials $V_j(x)$ and $V_{5j}(x)$ can only appear with an odd number of spatial derivatives. Due to the presence of spatial derivatives, they are less relevant (in the RG sense) compared to the mass and axial chemical potential disorders, and we have not included them in $H_{D,1}$. We can choose $V_4(x)$, $V_{45}(x)$ as independent random variables (as specified below). The disorder coupling constants will be denoted by $\Delta_a$ with $a=4, 45$. 

For class AIII, the disorder potentials have to anticommute with $\Gamma_4$ matrix (defining criterion of discrete chiral symmetry for this class). Therefore, we expect the disorder Hamiltonian for class AIII to be similar to the one for class DIII, i.e., we just need to exchange $\Gamma_4$ and $\Gamma_5$ matrices. However, in class AIII we are dealing with complex fermions rather than the real Majorana fermions realized in class DIII. In addition, the lack of time reversal symmetry for class AIII (or a remnant U(1) spin rotational symmetry paired states belonging to class AIII) also allows for a random abelian vector potential. Therefore, the general disorder Hamiltonian for class AIII has the form
\begin{eqnarray}\label{disorderAIII}
H_{D,2}= \int d^3x \; \Psi^\dagger(x) \; [V_5(\mathbf{x}) \Gamma_5 + V_{45}(\mathbf{x})i\Gamma_4 \Gamma_5 
\nonumber \\ + i \sum_{j=1}^{3} \; V_{4j}(x) \Gamma_4 \Gamma_j  +  \sum_{j=1}^{3} \; V_j(x) \Gamma_j ]\Psi(x), 
\end{eqnarray}
where $V_5$, $V_{45}$, $V_j$ and $V_{4j}$ respectively denote random mass, random axial chemical potential, random Abelian vector potential and a random spin orbit potential. We have ignored the possible disorder potentials which can be written with a spatial derivative operating on $\Psi$ as they are less relevant in the RG sense. For class CI and CII we again we have similar models of particle-hole symmetric disorder but involving eight component Dirac fermions. We again emphasize that in classes DIII, AIII, CI and CII for preserving particle-hole symmetry, we are not allowed to consider the conventional form of scalar potential disorder that couples to the number density $\Psi^\dagger \Psi$ of the Dirac fermions. Since class CI describes a time reversal invariant, spin singlet superconductor, it has global SU(2) spin rotational symmetry. Therefore, class CI can also allow for a random nonAbelian SU(2) vector potential (random spin gauge field) in addition to the mass and axial chemical potential disorders.

Since there is no stringent requirement of particle-hole symmetry for class AII, any time reversal symmetry preserving disorder potential is allowed for this symmetry class. As discussed in Ref.~\onlinecite{GoswamiChakravarty}, the Hamiltonian for generic time reversal symmetric disorder in class AII can be written as
\begin{eqnarray}\label{disorderAII}
H_{D,3}&=&\int d^3x \; \chi^\dagger(x) \; [V_0 (\mathbf{x}) \mathbb 1+ V_4 (\mathbf{x}) \Gamma_4 
+V_{45}(\mathbf{x})i\Gamma_4 \Gamma_5 \nonumber \\ &&+  i \sum_{j=1}^{3} \; V_{4j}(x) \Gamma_4 \Gamma_j ]\chi(x),
\end{eqnarray} 
where $V_0(\mathbf{x})$ $V_4 (\mathbf{x})$, $V_{45}(\mathbf{x})$ and $V_{4j} (\mathbf{x})$ respectively denote a random chemical potential, a random Dirac mass or band gap, a random axial potential, and a random spin orbit potential. We have not included $\sum_{j=1}^{3} \; f_{j,1}(\mathbf{x}) \chi^\dagger \Gamma_5 \Gamma_j \partial_j \chi$ and $\sum_{j=1}^{3} \; f_{j,2}(\mathbf{x}) \chi^\dagger \Gamma_4 \Gamma_5 \Gamma_j  \partial_j \chi$ as they are less relevant in the RG sense due to the presence of spatial derivatives.

\subsection{Generalized Harris criterion} When the disorder potentials follow Gaussian white noise distribution, under the scale transformations $x \to x e^l$, $\tau \to \tau e^{zl}$, the random potentials and the corresponding coupling constants respectively transform as
\begin{equation}
V_a \to V_a e^{l \eta_a} \implies \Delta_a \to \Delta_a e^{(2\eta_a-d)l}
\end{equation} where $\eta_a$ is the scaling dimension of the associated fermion bilinear. The disorder coupling $\Delta_a$ is a relevant (irrelevant) perturbation when
$$\eta_a > d/2 \:  (\eta_a<d/2).$$ It is a spectacular property of the noninteracting Dirac fixed point that the anomalous dimension of any symmetry allowed bilinear $\Psi^\dagger \hat{M}_a \Psi$ is equal to the dynamical exponent $z=1$. Therefore for the class DIII model, $\eta_{4}=\eta_{45}=1$, $\eta_{j5}=\eta_j=0$ (as the corresponding disorder potentials $V_{j5}$ and $V_j$ are accompanied by an additional spatial derivative). Therefore, all symmetry allowed disorder couplings are irrelevant perturbations, and the massless Dirac fixed point can survive against sufficiently weak disorder. Given that $\Delta_j$ and $\Delta_{j5}$ are more irrelevant than $\Delta_4$ and $\Delta_{45}$ (due to smaller scaling dimensions), in the following discussions we will mainly focus on the random mass and random axial potentials for class DIII and AIII problems. For class AII we also have identical stability criterion. The equality $\eta_a=1$ for the massless Dirac fixed point also shows that the effects of Gaussian white noise disorder can be studied by using a $d=2+\epsilon$ expansion. 

We can also formulate the stability criterion for a more general disorder distribution function and develop an alternative method of $\epsilon$ expansion for fixed spatial dimensions. For the Gaussian white noise distribution, the correlation between the disorder potentials at different spatial points is described by a Dirac delta function, which is a simple example of a homogeneous tempered distribution. Following the theory of harmonic functions, we can consider a more generalized homogeneous tempered distribution in terms of the Riesz potential~\cite{Stein}, which as a special limit reproduces the Dirac delta distribution. The Riesz potential is defined as
\begin{equation}
I_\alpha(\mathbf{x})=\frac{\Gamma\left(\frac{d-\alpha}{2}\right)}{2^\alpha \pi^{d/2} \Gamma(\alpha/2)} \; \frac{1}{|\mathbf{x}|^{(d-\alpha)}}
\end{equation}
and its Fourier transform is given by $|\mathbf{q}|^{-\alpha}$. Notice that for $\alpha=(d-2)$ we have the marginal case of $1/r^2$ power law correlations. Since
\begin{equation}
 \lim_{\alpha \to 0} \; \int d^d y f(\mathbf{y}) I_\alpha(\mathbf{x}-\mathbf{y})= f(\mathbf{x}),
\end{equation}
in terms of the singular Riesz potential, the Dirac delta distribution can be written as
 \begin{equation}
\delta^d(\mathbf{x}-\mathbf{y})=\lim_{\alpha \to 0} \frac{\Gamma\left(\frac{d-\alpha}{2}\right)}{2^\alpha \pi^{d/2} \Gamma(\alpha/2)} \; \frac{1}{|\mathbf{x}-\mathbf{y}|^{(d-\alpha)}}.
\end{equation}
Motivated by this we consider the disorder potentials with the following power law correlations
\begin{equation}
\langle V_j(\mathbf{x}) V_j(\mathbf{y}) \rangle =\Delta_j \; \frac{\Gamma\left(\frac{d-\alpha}{2}\right)}{2^\alpha \pi^{d/2} \Gamma(\alpha/2)} \;  \frac{1}{|\mathbf{x}-\mathbf{y}|^{(d-\alpha)}},
\end{equation}
and we are mainly interested in $d > 2$. When $\alpha=0$, we recover the Gaussian white noise distribution. For any $\alpha>0$, disorder potential has long range correlations. Under the scale transformations $x \to x e^l$, the disorder coupling transforms as
$$\Delta_j (l) =\Delta_0 e^{(2\eta_j +\alpha -d)l}.$$ Therefore, at the massless Dirac fixed point with $\eta_j=z=1$, $(2+\alpha-d)=-\epsilon$ determines the relevance ($\epsilon<0$) or irrelevance ($\epsilon >0$) of the disorder potential. Therefore, the perturbative calculations can be controlled by a small $\epsilon$ expansion. 

Notice that we have two possible ways for carrying out such a calculation: (I) by holding the spatial dimension fixed, we can continue in the parameter $\alpha$ as $\alpha=\alpha_m-\epsilon$, since $\alpha=\alpha_m=(d-2)$ corresponds to the marginal case; (II) by holding $\alpha$ fixed, we can continue in the number of spatial dimensions as $d=d_m+\epsilon$ with $d_m=\alpha+2$. The  method (I) will be used in this manuscript for studying general case of long range power law correlated disorder and topological transitions at fixed dimensions, and it is similar to the scheme used by Nayak and Wilczek for clean interacting fermions~\cite{Nayak}. The method (II) is a generalization of the $d=2+\epsilon$ expansion method conventionally used for the Gaussian white noise disorder. The cut-off procedures employed by these two schemes are not analytically related, and  $\beta$-functions in general are known to depend on the precise regularization scheme. Only the universal critical properties have to be independent of the  scheme; otherwise the fixed point itself has to be independent of the scheme, which is extremely rare. For example, even for the widely studied case of $O(N)$ non-linear $\sigma$-model {\em only} terms up to two loops in the $\beta$-function in $d=2$ are universal, as long as the regularization schemes are analytically related~\cite{Creutz}. Even though we have formulated the generalization of Harris criterion by keeping a quantum critical point (repulsive fixed point) in mind, the methodology is equally applicable for describing the stability of an attractive fixed point.

\section{Nonperturbative rare events and localization for weak disorder}\label{sec:rare}
Before considering the effects of stronger disorder on the quantum critical behavior, it is important to ask about the nonperturbative effects of disorder on the clean, gapped states and the direct topological quantum phase transition. The perturbative calculations do suggest that the density of states for any energy below the average mass gap vanishes for sufficiently weak disorder. However, nonperturbative effects due to statistically rare events (corresponding to large fluctuations) can destroy such sharp spectral gaps by inducing density of states below the disorder averaged mass gap. Unless these states are localized, we cannot associate any sharp notion of topologically distinct insulating states. Therefore, it is best to consider the massless Dirac fixed point as the quantum critical point between topologically distinct localized states. Here we will derive simple formulas for the low energy density of states and the localization length for weak particle hole symmetric disorder. For concreteness we will consider class DIII model. 

We note that the effects of nonperturbative rare events due to random scalar potential disorder on a Dirac and Weyl semimetal have been originally discussed in Ref.~\onlinecite{Nandkishore} and subsequently have been numerically verified in Ref.~\onlinecite{Pixley3}. They show how scalar potential induced quasi-localized rare states with power law wavefunctions can give rise to a finite density of states at zero energy. This turns the perturbatively found scalar potential disorder controlled critical point (governing the semimetal-metal transition) into an avoided critical point. However, the large quantum critical fan predicted by perturbative analysis survives up to a very low energy or long length scale~\cite{Pixley3}, in contrast to the original anticipation of Ref.~\onlinecite{Nandkishore}. Whether such rare states can destroy the direct topological QPT between two insulating states in class AII (in the presence of generic time reversal symmetry preserving disorder) is still an open problem, since the models studied so far are only concerned with the stability of a semimetal phase, rather than topological quantum criticality. \emph{Here we will be addressing the effects of exponentially localized rare states introduced by particle-hole symmetric disorder on the clean, gapped phases, and the methods of Ref.~\onlinecite{Nandkishore} are not directly applicable to the generic problem of particle-hole symmetric disorder}. We also note that for conventional Wigner-Dyson symmetry classes (A, AI, AII), Wegner has showed that the Gaussian disorder generally induces a finite density of states~\cite{Wegner81}. But no such general result is known for other six particle-hole symmetric AZ symmetry classes.

The rare events are accessed by studying off critical properties, by keeping a fixed disorder averaged mass $|\mu|$ such that $\xi_M= \hbar v/|\mu|$ acts as the shortest length scale. Therefore, $|\mu|> \mathrm{max} \{ k_B T, \hbar \omega, \hbar v/L \}$, where $\omega$ is the frequency, $L$ is the system size, and $T$ is the temperature. For a superconducting system the band parameters can be functions of temperature. For simplicity let us consider the mass disorder. Imagine that a statistically rare event gives rise to a compact TSC region of linear dimensions $R$, within the NSC phase with an average $\bar{\mu}<0$. For simplicity, we will consider this region to be a sphere of radius $R$. Within this TSC region, the mass disorder potential satisfies $\bar{\mu}<V_4<\infty$. Due to the change in the topological invariant, such a domain supports low energy bound/surface states when $E< \mathrm{min} \{\bar{\mu}, \bar{\mu}+V_4 \}$. However, the energy of such surface state will be proportional to the inverse of the radius of compactification.  The energy eigenvalues of such bound states can be obtained by solving the Dirac equation for a spherical potential. The even parity ($+$) solutions have total angular momentum $j=l+1/2$, $l=0,1,..$, and $m=-j, -j+1, ..., j-1, j$, while the odd parity ($-$) solutions correspond to $j=l-1/2$ and $l>0$. These solutions can be written as
\begin{equation}
\psi^{\pm}_{j,m}=\begin{bmatrix}
\frac{i G_{lj}}{r} \varphi^{\pm}_{jm}(\theta,\phi) \\
\frac{F_{lj}}{r} \varphi^{\mp}_{jm}(\theta, \phi)
\end{bmatrix},
\end{equation} where we have introduced two component spinor harmonics
\begin{equation}
\varphi^{\pm}_{jm}(\theta,\phi)=\frac{1}{\sqrt{2l+1}}\begin{bmatrix}
\sqrt{l\pm m+1/2} \; Y^{m-1/2}_{l}(\theta,\phi) \\
\pm \sqrt{l \mp m+1/2} \; Y^{m+1/2}_{l}(\theta,\phi)
\end{bmatrix}.
\end{equation}  As mentioned above, we will take the mass profile to be $\mu(r)=\mu_1 \Theta(R-r)-\mu_2 \Theta(r-R)$, where $\Theta(r)$ is the Heaviside step function, $\mu_2 = \bar{\mu}>0$, and $0<\mu_1<\infty$. After introducing the notations $\kappa_1=\sqrt{\mu^2_1-E^2}/(\hbar v)$ and $\kappa_2=\sqrt{\mu^2_2-E^2}/(\hbar v)$, the bound state solutions for $r<R$ are written in terms of modified spherical Bessel functions of the first kind, while those in the region $r>R$ are expressed in terms of modified Bessel functions of second kind. For $j=l \pm 1/2$, the radial functions are given by
\begin{eqnarray}
G_{lj}&=&r[i_{l}(\kappa_1 r) \Theta(R-r)+ k_{l} (\kappa_2 r) \Theta (r-R)], \\
F_{lj}&=&r \bigg[i_{l \pm 1}(\kappa_1 r) \frac{\kappa_1}{E-\mu_1} \Theta(R-r) - k_{l \pm 1}(\kappa_2 r) \nonumber \\ && \times  \frac{\kappa_2}{E+\mu_2} \Theta(r-R) \bigg].
\end{eqnarray} After matching the interior and the exterior solutions we find the following conditions
\begin{eqnarray}
\frac{i_{l}(\kappa_1R) k_{l \pm 1}(\kappa_2 R)}{i_{l\pm 1}(\kappa_1R) k_{l}(\kappa_2 R)}=\sqrt{\frac{(\mu_1+E)(\mu_2+E)}{(\mu_1-E)(\mu_2-E)}},
\end{eqnarray}where $\pm$ signs respectively correspond to $j=l \pm 1/2$. For $|E|$ much smaller than $\mu_1$ and $\mu_2$, we can safely use the asymptotic expansions for the Bessel functions. After some simple algebra we find that $j=l+1/2$ solutions (with $l=0,1,2...$) can only give positive eigenvalues $E \approx (l+1) \hbar v/R=(j+1/2)\hbar v/R$, while $j=l-1/2$ solutions (with $l=1,2...$) lead to negative eigenvalues $E \approx - l \hbar v/R=-(j+1/2)\hbar v/R$. Therefore, the low energy spectra are determined by
\begin{equation}
E=\pm (j+1/2) \frac{\hbar v}{R} \Theta(R-\xi_M)
\end{equation}
and each level has a degeneracy $(2j+1)$. 

For simplicity if we assume Gaussian white noise distribution for mass disorder, the effects of low energy bound states on the average density of states can be estimated in the following manner. The probability for finding the NSC domain of size $R$ inside the bulk TSC is roughly given by $P[V_4] \sim \exp[-c R^d]$, with $c=V^2_4 /(2\Delta_4)$. Therefore, the low energy density of states will be 
\begin{eqnarray}
\rho(E) & \sim & \int^{\infty}_{\bar{\mu}} dV_4 \; \int^{\infty}_{\xi_M} dR \; \exp \left[-c R^d\right]  \; \delta \left(|E|- \frac{\hbar v}{R} \right) \nonumber \\ & \sim & \int^{\infty}_{\bar{\mu}} dV_4 \; |E|^{-2} \exp \left(-c \frac{(\hbar v)^d}{|E|^d}\right) \; , \nonumber \\ &\sim & |E|^{(d/2-2)} \; \mathrm{erfc}\left(\frac{\bar{\mu}(\hbar v)^d/2}{\Delta_4 |E|^{d/2}}\right)\label{LTailDOS}
 \end{eqnarray} where $\mathrm{erfc}(x)$ is the complementary error function. For $|E| \ll \bar{\mu}$, the argument of $\mathrm{erfc}(x)$ is very large and the asymptotic behavior $\mathrm{erfc}(x) \sim \exp(-x^2)/\sqrt{\pi}x$ leads to 
\begin{equation}
\rho(E) \sim |E|^{(d-2)} \; \exp \left(-\frac{\bar{\mu}^2 (\hbar v)^d}{\Delta_4 |E|^d}\right).
\end{equation}
From this estimation we see that the rare events give rise to a Lifshitz tail in the average density of states and destroy the hard spectral gap of the NSC and TSC phases. The prefactor $|E|^{(d-2)}$ is capturing the density of states due to surface bound states localized on the rare region (corresponds to the $|E|$ linear density of states of two dimensional massless Dirac/Majorana fermion bound at the interface between two topologically distinct states), while $\exp(-c^\prime |E|^{-d})$ part is accounting for its small probability. These considerations are not qualitatively modified by the presence of a random axial chemical potential. For $d=3$, the asymptotic behavior of the low energy density of states inside the insulating phase is given by $\log(\rho(E)) \sim - |E|^{-3}$, reflecting that the density of states vanishes at zero energy. These Griffith singularities affect the thermodynamic properties of  the localized phase at low temperatures in a very weak manner. Since they do not cause any divergent density of states, they cannot lead to activated dynamic scaling found for one dimensional systems [one finds $|E| \sim \exp(-c^{\prime \prime} L^d)$ and $d \bar{\nu}<2$ for one dimensional problems]. Therefore, we do not expect these weak Griffith singularities to destroy the quantum critical properties of massless Dirac fermions in class DIII. We note that the property $\log(\rho(E)) \sim - |E|^{-2}$ inside the two dimensional insulating phase has been observed in the numerical work on two dimensional model belonging to class D~\cite{Mirlin}. It is also interesting to note that the quantum phase transition between topologically distinct localized states in class D was also shown to be governed by the two dimensional, massless Majorana fermion fixed point, despite the presence of rare states. 

To summarize, the nonperturbative rare regions for particle-hole symmetric disorder cause a finite density of states below the average mass gap, which however vanishes at zero energy. The localization length for the disorder induced low energy states is given by the inverse of average Dirac mass or $\xi_M$. When the disorder strength is gradually increased, these bound states become more abundant, and tunneling between the localized states can ultimately give rise to a diffusive phase with finite density of states at zero energy. However this is a difficult problem to tackle analytically. Hence, in the subsequent sections we will restrict ourselves with disorder averaged perturbative analysis.

\begin{figure}[htb]
\centering
\includegraphics[scale=0.6]{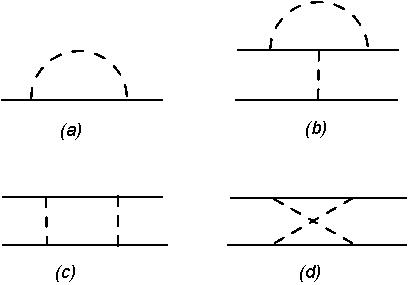}
\label{fig:4a}
\caption{Feynmann diagrams arising from disorder averaged quartic interactions. (a) disorder induced fermion self energy, (b) disorder vertex correction, (c) ladder and (d) crossing type diagrams. At one loop level the diagrams of class (a) and (b) are UV  divergent for any $\alpha>0$ and contribute to the beta function. By contrast, the diagrams of class (c) and (d) at one loop level are UV convergent for all $\alpha>0$ and do not contribute to the RG flow equations of the disorder couplings. Therefore the one loop beta function of a disorder vertex for any $\alpha>0$ is determined by the anomalous scaling dimension of an appropriate fermion bilinear. For example, if we consider a disorder $V_j(\mathbf{x}) \; \Psi^\dagger \hat{M}_j \Psi$, where $\hat{M}_j$ is a $4 \times 4$ matrix, the one loop beta function of the disorder coupling will be determined by the anomalous dimension $\eta_j$. For $\alpha=0$, both classes of diagrams are UV divergent and can contribute to the beta function as found within $d=2+\epsilon$ expansion scheme}.
\end{figure}

\section{Localization-delocalization transitions at stronger disorder} \label{sec:LD} 
The breakdown of superuniversality or massless Dirac fixed point for sufficiently strong disorder introduces an intervening diffusive metal phase between two topologically distinct localized phases. Therefore, beyond a critical strength of disorder no direct transition between two localized states is expected, allowing transitions between the diffusive phase and two localized states. We address the disorder driven breakdown of superuniversality by using perturbative renormalization group analysis. In this section we will mainly work with long range disorder potentials with $\alpha>0$, employing the scheme (I). 

We first elucidate some important technical aspects of perturbative renormalization group calculation. Consider the one loop vertex correction diagram of Fig.~3(b), which involves the integral
$$\int \frac{d^dq}{(2\pi)^d} \hat{M}_j G(\mathbf{k}-\mathbf{q},0) \hat{M}_j G(\mathbf{k}-\mathbf{q},0) \hat{M} D_j(\mathbf{q}),$$ where $D_j(\mathbf{q})=\Delta_j |\mathbf{q}|^{-\alpha}$ is the disorder propagator. We are interested in the part $|\mathbf{q}| \gg |\mathbf{k}|$. After setting $k=0$, this integral depends on the UV cut-off as 
$$\int \frac{1}{k^2} \; \frac{1}{k^\alpha} \; k^{(d-1)} dk \sim \Lambda^{(d-2-\alpha)}.$$ Within both schemes (I) and (II), this integral behaves as $\sim \Lambda^{\epsilon}$. Same behavior is found for self-energy diagram of Fig.~3(a).

Now consider the one loop ladder and crossing type diagrams of Fig.~3(c) and 3(d). These diagrams involve the integral
$$\int \frac{d^dq}{(2\pi)^d} [\hat{M}_j G(\mathbf{k}-\mathbf{q},0) \hat{M}_j][\hat{M}_j G(\mathbf{k} \pm \mathbf{q},0) \hat{M}_j]D^2_j(\mathbf{q}).$$ When external leg momentum is much smaller than $|\mathbf{q}|$, these integrals depend on the UV cut-off as
$$\int \frac{1}{k^2} \; \frac{1}{k^{2\alpha}} \; k^{(d-1)} dk \sim \Lambda^{(d-2-2\alpha)}.$$ If we hold $d$ to be fixed (scheme (I)) and continue in $\alpha$, the integrals behave as $\Lambda^{(2-d)} \Lambda^{2\epsilon}$ (setting $\alpha=d-2-\epsilon$). For any $d>2$ and a very small $\epsilon$ (as required for any expansion scheme), the integrals are UV-convergent. On the other hand, if we hold $\alpha$ fixed (scheme (II)) and continue in $d$, the integrals behave as $\Lambda^{d-2-\alpha}\Lambda^{-\alpha}=\Lambda^\epsilon\Lambda^{-\alpha} $. For any $\alpha >0$ and a very small $\epsilon$, the integrals are again UV-convergent. Therefore, for $\alpha>0$ and $d>2$, we can carry out a calculation using either methods where ladder and crossing diagrams are suppressed.

Notice that by directly setting $\alpha=0$ will make all the integrals to behave as $\Lambda^\epsilon$ (as found within $2+\epsilon$ expansion scheme). Whether the physical answers (universal critical properties) obtained by taking $\alpha \to 0$ limit of our calculations will indeed reproduce the results of a direct $d=2+\epsilon$ expansion is not entirely clear. Therefore, we will make the above arguments more formal, and show that the one loop beta functions obtained within two schemes are different, but dynamical scaling and correlation length exponents for any given $\alpha$ are the same. The main difference between the two types of leading order calculations arise regarding the predictions for the anomalous scaling dimension of the fermion field. For simplicity we first consider the effects of scalar potential type disorder with $\hat{M}=\mathbb 1$.

We consider the field theoretic evaluation of the integral
\begin{equation}
I_1=\int_q \frac{1}{(\mathbf{k}+\mathbf{q})^2} \; \frac{1}{|\mathbf{q}|^{\alpha}}
\end{equation}
that arises for the vertex diagram. After using the Feynmann parameter we find
\begin{eqnarray}
I_1&=&\int^{1}_{0} dx \; \int_q \; \frac{\alpha}{2} \frac{(1-x)^{\alpha/2-1}}{[(\mathbf{q}+x\mathbf{k})^2+x(1-x)k^2]^{\alpha/2+1}} \nonumber \\
&=& \frac{\alpha}{2} \frac{k^{(d-\alpha-2)}}{(4\pi)^{d/2}} \frac{\Gamma(\alpha/2+1-d/2)\Gamma(d/2-\alpha/2)\Gamma(d/2-1)}{\Gamma(d-\alpha/2-1)\Gamma(1+\alpha/2)} \nonumber \\
\end{eqnarray}
If we set $\alpha=(d-2)-\epsilon$, the UV divergent part becomes 
\begin{equation}
I_1 \sim \frac{1}{(4\pi)^{d/2}} \frac{1}{\Gamma(d/2)} \left(\frac{-2}{\epsilon}\right).
\end{equation}
If we set $d=2+\alpha+\epsilon$, we again find
\begin{equation}
I_1 \sim \frac{1}{(4\pi)^{1+\alpha/2}} \frac{1}{\Gamma(1+\alpha/2)} \left(\frac{-2}{\epsilon}\right),
\end{equation}
confirming that within both methods, (I) and (II), $I(k=0)\sim \Lambda^\epsilon$.

Now, we estimate the integral
\begin{equation}
I_2=\int_q \frac{1}{(\mathbf{k}+\mathbf{q})^2} \; \frac{1}{|\mathbf{q}|^{2\alpha}}
\end{equation}
which arises for the ladder or crossing type diagrams. We find
\begin{equation}
I_2=\alpha \frac{k^{(d-2\alpha-2)}}{(4\pi)^{d/2}} \frac{\Gamma(\alpha+1-d/2)\Gamma(d/2-\alpha)\Gamma(d/2-1)}{\Gamma(d-\alpha-1)\Gamma(1+\alpha/2)}
\end{equation}
Within the  method (I) of continuing in $\alpha$, 
\begin{equation}
I_2 \sim \frac{(d-2)}{(4\pi)^{d/2}} \frac{\Gamma^2(d/2-1) \Gamma(2-d/2)}{\Gamma(d-1)} k^{(2-d)}
\end{equation}
which is indeed UV-convergent for $d=3$ and small $\epsilon$. Within the dimensional continuation scheme 
\begin{equation}
I_2 \sim \frac{\alpha}{(4\pi)^{1+\alpha/2}} \frac{\Gamma^2(\alpha/2) \Gamma(1-\alpha/2)}{\Gamma(\alpha+1)} k^{-\alpha}
\end{equation}
which is UV-convergent for $\alpha>0$ and small $\epsilon$.

The one loop fermion self energy diagram of Fig.~3(a) due to random scalar potential is found to be
\begin{widetext}
\begin{eqnarray}
\Sigma \sim -\Delta_0 \; \frac{\alpha}{2} \; \frac{1}{(4\pi)^{d/2}} \; \frac{\Gamma(\alpha/2+1-d/2)\Gamma(d/2-\alpha/2)\Gamma(d/2-1)}{\Gamma(d-\alpha/2-1)\Gamma(1+\alpha/2)}\left[i\omega + \mathbf{k}\cdot \boldsymbol \Gamma \frac{(d-2)}{2d-\alpha-2}\right] \label{eqnimp}
\end{eqnarray}
\end{widetext}
After identifying $1/\epsilon$ with $\log \Lambda=-l$, the RG flow equation within the  scheme (I) becomes
\begin{equation}
\frac{d\Delta_0}{dl}=(2z-d+\alpha)\Delta_0=[-\epsilon+4\Delta_0(1-1/d)]\Delta_0
\end{equation}
where we have redefined $\frac{2 \Delta_0}{\Gamma(d/2) (4\pi)^{d/2}} \to \Delta_0$, $z(l)=1+2\Delta_0(l)(1-1/d)$. The flow equation within scheme (II) has the form
\begin{equation}
\frac{d\Delta_0}{dl}=(2z-d+\alpha)\Delta_0=\left[-\epsilon+4\Delta_0 \frac{\alpha+1}{\alpha+2}\right]\Delta_0
\end{equation}
where we have redefined $\frac{2\Delta_0}{(4\pi)^{1+\alpha/2}} \frac{1}{\Gamma(1+\alpha/2)} \to \Delta_0$ and $z(l)=1+2\Delta_0(l)\frac{\alpha+1}{\alpha+2}$. At the disorder controlled critical point both methods give $z=1+\epsilon/2=(d-\alpha)/2$ and $\nu=1/\epsilon=1/(d-2-\alpha)$. If we take $\alpha \to 0$, we do recover the one loop result $z=3/2$, $\nu=1$ for Gaussian white noise disorder, obtained earlier within $d=2+\epsilon$ expansion scheme. We expect that for $\alpha>0$, the beta function within both schemes will have the form $[2z(\Delta_0)-d+\alpha]\Delta_0$ even at higher loops, if the ladder and crossing type diagrams remain suppressed (as these diagrams always possess more disorder propagators in comparison to the vertex correction or self energy diagrams).

The main difference between the predictions from these two schemes arises from the different coefficients obtained for $\mathbf{k}\cdot \boldsymbol \Gamma$ in Eq.~(\ref{eqnimp}). Within the scheme (I), the coefficient of $\mathbf{k}\cdot \boldsymbol \Gamma$ inside the parenthesis will be estimated as $(d-2)/(d+\epsilon) \approx (d-2)/d$. On the other hand, the same coefficient within the scheme (II) will become $(\alpha + \epsilon)/(\alpha+2+2\epsilon) \approx \alpha/(\alpha+2)$. Consequently, the anomalous dimension of the fermion field within the scheme (I) is found to be $\eta_\Psi=z-1-(d-2)\epsilon/(d-1)=d \epsilon/[4(d-1)]=d(1-\alpha)/[4(d-1)]$. This should be contrasted with $\eta_\Psi=\epsilon (\alpha +2)/[4(\alpha +1)]=(d-2)(\alpha +2)/[4(\alpha +1)]$ found by using the  scheme (II). If we take the formal limit $\alpha \to 0$ for Gaussian white noise distribution, we respectively find $\eta_\Psi=d(d-2)/[4(d-1)]$ and $\eta_\Psi=(d-2)/2$ from schemes (I) and (II). We have already stressed that which expansion method at the leading order works better for Gaussian white noise disorder is not a priori clear. Comparison with numerical calculations can provide  further insight into this delicate issue. After solving the RG flow equations, we can show that the quasiparticle residue of massless Dirac fermion (at zero frequency) vanishes as
\begin{equation}
Z(\mathbf{k}) \approx \left[\delta +(1-\delta) \left(|\mathbf{k}|/\Lambda\right)^{1/\nu} \right]^{\eta_\Psi \nu_\Delta},
\end{equation}
when the critical point is approached from the semimetal side. At the same time, the effective Fermi velocity vanishes as 
\begin{equation}
v(\mathbf{k}) \approx v \left[\delta +(1-\delta) \left(|\mathbf{k}|/\Lambda\right)^{1/\nu} \right]^{(z-1) \nu_\Delta}.
\end{equation}
Therefore, inside the semimetal regime defined by $\delta >> (1-\delta) \left(|\mathbf{k}|/\Lambda\right)^{1/\nu}$, the residue vanishes as $\delta^{3/8}$ and $\delta^{1/2}$ respectively within schemes (I) and (II) for Gaussian white noise disorder. Since both methods predict the same $z$, the Fermi velocity vanishes as $\delta^{1/2}$ within both schemes. In the critical regime $\delta >> (1-\delta) \left(|\mathbf{k}|/\Lambda\right)^{1/\nu}$, $Z(K) \sim k^{\eta_\Psi}$ and $v(k) \sim k^{(z-1)}$. Therefore, the critical fermion propagator at zero frequency behaves as 
\begin{eqnarray}
G(\omega=0,\mathbf{k}) & \sim & \frac{Z(\mathbf{k})}{v(\mathbf{k}) \mathbf{k} \cdot \boldsymbol \Gamma} \sim  \frac{\hat{k} \cdot \boldsymbol \Gamma}{v k^{(z-\eta_\Psi)}}.
\end{eqnarray}
For Gaussian white noise, the scheme (I) leads to $G \sim k^{-9/8}$, while the scheme (II) yields $G \sim k^{-1}$ just like the propagator of clean Dirac fermions. Therefore, numerical studies of $G$ will be important to understand which method actually works better for Gaussian white noise disorder. 

We can similarly consider the effects of mass disorder for $\alpha>0$. The anomalous dimension of the mass operator $\eta_4$ is a complicated unknown function of the disorder coupling $\Delta_4$. When the ladder and crossing type diagrams are suppressed, the beta function for mass disorder will have the form 
\begin{equation}
\frac{d\Delta_4}{dl}=\Delta_4[2\eta_4(\Delta_4)-d+\alpha].
\end{equation} If a nontrivial zero of the beta function exists, then quite generally it will require $\eta_4(\Delta^\ast_4)=(d-\alpha)/2$. Since the localization length exponent $\nu_M=1/\eta_4$, any nontrivial zero of the beta function will satisfy $\nu_M=2/(d-\alpha)$. 

Next we discuss the effects of a random axial potential. The beta function for dimensionless axial disorder coupling has the form
\begin{equation}
\frac{d\Delta_{45}}{dl}=\Delta_{45}[2\eta_{45}(\Delta_{45})-d+\alpha].
\end{equation}
Due to the properties $[\Gamma_j,\Gamma_4\Gamma_5]=0$, $\{ \Gamma_j, \Gamma_4 \}=0$ and $\{ \Gamma_4, \Gamma_4 \Gamma_5 \}=0$, we can show that $\eta_{45}(\Delta_{45})=\eta_{4}(\Delta_{45})=\eta_1(\Delta_{45})$~[see Appendix~\ref{sec:AppendixA}]. Consequently, at a nontrivial fixed point controlled by the axial disorder, we have $\eta_{45}(\Delta^\ast_{45})=\eta_{4}(\Delta^\ast_{45})$, implying the relation $\nu_M=1/z=2/(d-\alpha)$. For this reason we anticipate that $\nu_M=2/(d-\alpha)$ to be valid for a generic combination of disorder within class DIII. Similar statements can be made for classes AIII, CI and CII, which can only possess particle-hole symmetric disorders. Now we will perform explicit calculations for generic symmetry allowed disorder potentials for symmetry classes DIII, AII and AIII at one loop order by employing scheme (I).

\begin{figure*}[htb]
\centering
\subfigure[]{
\includegraphics[scale=0.45]{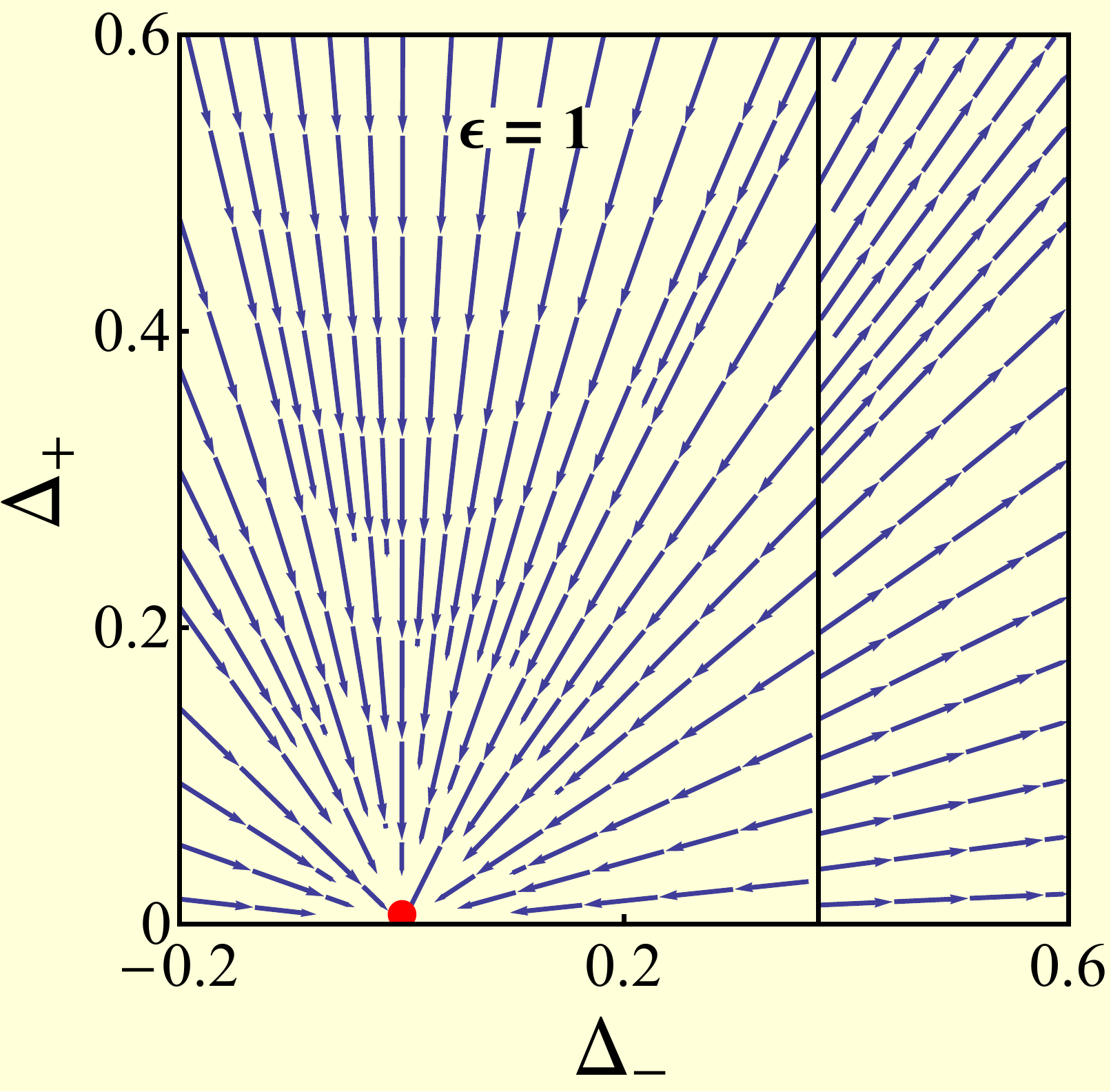}
\label{fig:5a}}
\subfigure[]{
\includegraphics[scale=0.45]{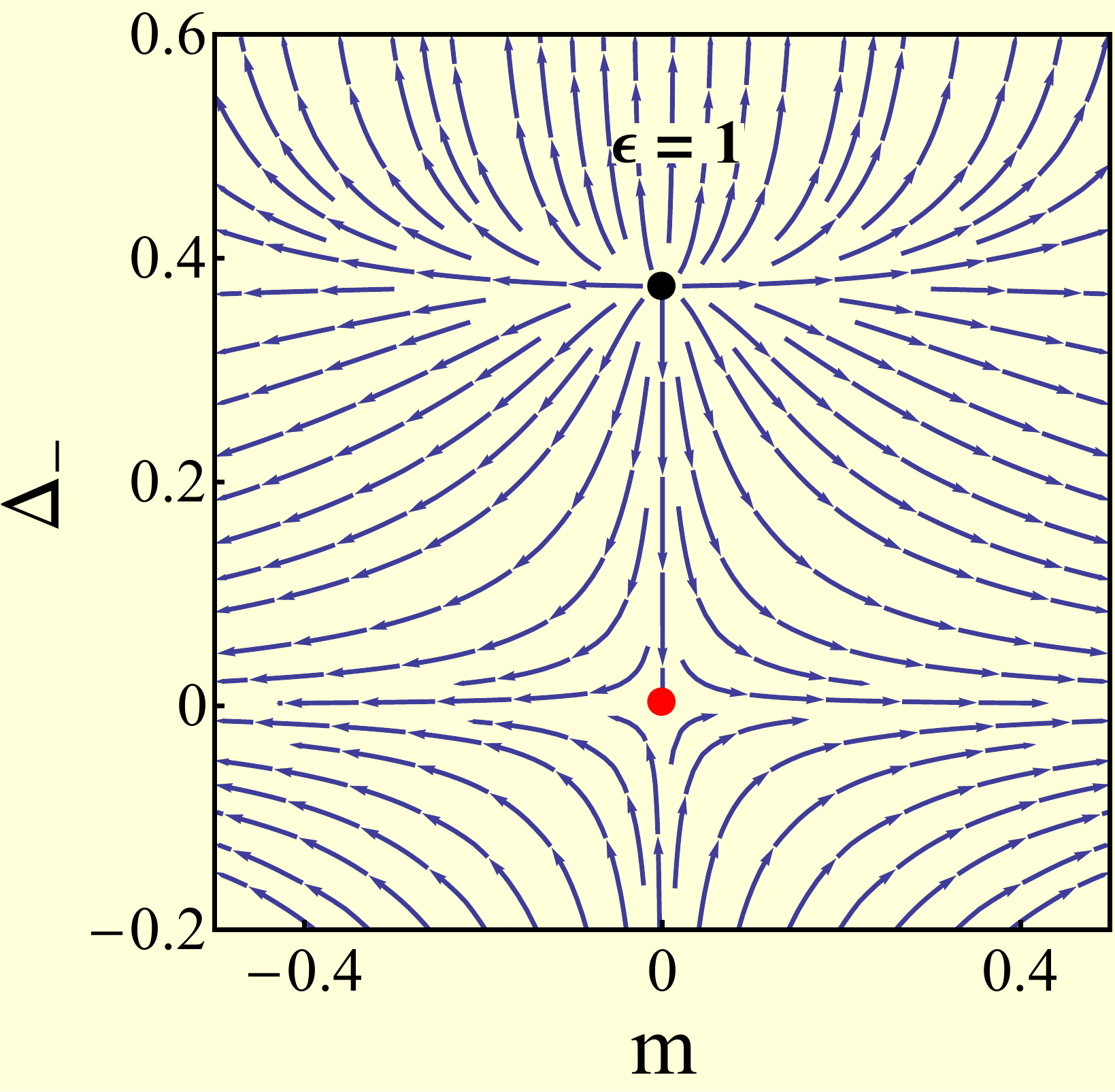}
\label{fig:5b}}
\caption{(Color Online) RG flows for class DIII (a) in the plane $\Delta_- -\Delta_+$ and (b) in the plane $m-\Delta_-$. We have chosen Gaussian white noise distributions for the mass disorder and axial disorder couplings respectively given by $\Delta_4$ and $\Delta_{45}$, and we also note that $\Delta_\pm=\Delta_{45} \pm \Delta_4$. (a) The solid vertical line at $\Delta^\ast_-=3/8$ describes the line of fixed points controlling the semimetal-metal and metal-insulator transitions (b) The black dot is a projection of the line of fixed points. The dot always describes the clean, massless Dirac fixed point. Our calculations suggest similar flow diagrams for class AIII.} 
\label{RGFlow}
\end{figure*}

\subsection{Class DIII at d=3}
The class DIII has two important disorder coupling constants $\Delta_4$ and $\Delta_{45}$ respectively corresponding to mass and axial chemical potentials [see Eq.~(\ref{disorderDIII})]. The one loop RG flow equations for these couplings and the Dirac mass are given by
\begin{eqnarray}
\frac{d\Delta_a}{dl}&=&\Delta_a \; \left[2\eta_a+\alpha-3 \right], \label{DIIIbeta} \\
\frac{dm}{dl}&=& m \; \eta_4 \; \label{DIIIbeta1}.
\end{eqnarray}
As shown in the Appendix~\ref{sec:AppendixB} the scaling dimensions are respectively given by
\begin{eqnarray}
\eta_4=\eta_{45}=1-\frac{4}{3}\Delta_4+\frac{4}{3}\Delta_{45}.\label{DIIIbeta2}
\end{eqnarray}
In the RG equations we are using dimensionless couplings $\Delta_a \Lambda^\epsilon /(2\pi^2 v^2)$. In addition, the field renormalization constant and the scaling dimension of the operator $\Psi^\dagger \Psi$ are respectively given by
\begin{eqnarray}
&&Z_{\Psi}=e^{-3l}[1-(\Delta_4+\Delta_{45})l],\label{DIIIbeta3}\\
&& \eta_{1}=z(l)=1+\frac{2}{3}\Delta_4 + \frac{4}{3}\Delta_{45}\label{DIIIbeta4},
\end{eqnarray}
where $z(l)$ is the scale dependent dynamic scaling exponent and $l=-\log \Lambda$. Two beta functions can also be rewritten in terms of new variables $\Delta_\pm=\Delta_{45} \pm \Delta_4$ as
\begin{eqnarray}
\frac{d \Delta_\pm}{dl}&=&\Delta_\pm \; \left[-\epsilon + \frac{8}{3}\Delta_-\right], \label{flowDIII}
\end{eqnarray}
with $\epsilon=1-\alpha$. The RG flows described by Eq.~(\ref{flowDIII}) are shown in Fig.~\ref{fig:5a}. We also demonstrate the flows involving Eq.~(\ref{DIIIbeta1}) and Eq.~(\ref{flowDIII}) in Fig.~\ref{fig:5b}. For any arbitrary value of $\Delta_+$, the zero of these two beta functions is determined by the line of fixed points
\begin{eqnarray}
\Delta^\ast_-=\Delta^\ast_{45}-\Delta^\ast_4=\frac{3}{8}(1-\alpha)=\frac{3\epsilon}{8},\label{DIIILFP}
\end{eqnarray}
for which the localization length exponent is given $\nu_M=1/\eta_4=2/(3-\alpha)$. Since the Gaussian white noise corresponds to $\alpha=0$ or $\epsilon=1$, we find $\nu_M=2/3$. However, the dynamic scaling exponent varies continuously along this line according to
$z^\ast=(3-\alpha)/2+ 2 \Delta^\ast_4$. On the other hand, the correlation length exponent for the disorder coupling is given by
$\nu_\Delta=1/\epsilon=1/(1-\alpha)$. Since the fermion field has a nontrivial anomalous dimension $\eta_\Psi=(\Delta^\ast_4+\Delta^\ast_{45})=3\epsilon/8+2\Delta^\ast_4$ (evaluated by setting $Z_{\Psi}=e^{-(3+\eta_\Psi)l}$), the quasiparticle residue of massless Dirac fermion vanishes according to $\delta^{\eta_\Psi \nu_\Delta}$ upon approaching the disorder controlled fixed points. We have denoted the reduced distance from the line of fixed points by $\delta=1-\Delta_-(0)/\Delta^\ast_-$, where $\Delta_-(0)$ is the bare value of $\Delta_-(l)$. While $\eta_M$ should remain $2/(3-\alpha)$ for all orders of perturbation theory, the other exponents $\eta_\Psi$, $z$ and $\nu_\Delta$ generically acquire corrections from higher order diagrams.

\subsection{Class AII at d=3} For class AII we have four coupling constants $\Delta_0$, $\Delta_4$, $\Delta_{45}$ and $\Delta_{4j}$ [see Eq.~(\ref{disorderAII})]. The RG flow equations for the corresponding dimensionless couplings at one loop order have similar forms as in Eq.~(\ref{DIIIbeta}) and the required scaling dimensions are given by
\begin{eqnarray}
\eta_1&=&z=1+\frac{4}{3}\Delta_0+\frac{2}{3}\Delta_4+\frac{4}{3}\Delta_{45}+\frac{10}{3}\Delta_{4j}, \label{AIIbeta1}\\
\eta_4&=&1-\frac{2}{3}\Delta_0-\frac{4}{3}\Delta_4+\frac{4}{3}\Delta_{45}+\frac{10}{3}\Delta_{4j}, \label{AIIbeta2}\\
\eta_{45}&=&1+\frac{4}{3}\Delta_0-\frac{4}{3}\Delta_4+\frac{4}{3}\Delta_{45}-\frac{8}{3}\Delta_{4j}, \label{AIIbeta3}\\
\eta_{4j}&=&1+\frac{2}{3}\Delta_0-\frac{2}{3}\Delta_4, \label{AIIbeta4}\\
Z_\Psi&=&e^{-3l}[1-(\Delta_0+\Delta_4+\Delta_{45}+3\Delta_{4j})l] \label{AIIbeta5}.
\end{eqnarray}
The explicit forms of the flow equations are
\begin{eqnarray}
&&\frac{d\Delta_0}{dl}=\Delta_0 \left[-\epsilon+\frac{8}{3}\Delta_0 +\frac{4}{3}\Delta_4+\frac{8}{3}\Delta_{45}+\frac{20}{3}\Delta_{4j}\right], \label{AIIbeta6} \nonumber \\ \\
&&\frac{d\Delta_4}{dl}=\Delta_4 \left[-\epsilon-\frac{4}{3}\Delta_0 -\frac{8}{3}\Delta_4+\frac{8}{3}\Delta_{45}+\frac{20}{3}\Delta_{4j}\right],\label{AIIbeta7}\nonumber \\ \\
&&\frac{d\Delta_{45}}{dl}=\Delta_{45} \left[-\epsilon+\frac{8}{3}\Delta_0 -\frac{8}{3}\Delta_4+\frac{8}{3}\Delta_{45}-\frac{16}{3}\Delta_{4j}\right],\label{AIIbeta8}\nonumber \\ \\
&&\frac{d\Delta_{4j}}{dl}=\Delta_{4j}\left[-\epsilon+\frac{4}{3}\Delta_0-\frac{4}{3}\Delta_4\right], \label{AIIbeta9} \\
&&\frac{dm}{dl}=m\left[1-\frac{2}{3}\Delta_0-\frac{4}{3}\Delta_4+\frac{4}{3}\Delta_{45}+\frac{10}{3}\Delta_{4j}\right].\label{AIIbeta10} \nonumber \\ 
\end{eqnarray} 
Notice that we can enforce the condition of discrete particle-hole symmetry of class DIII by setting the coupling constants of random scalar potential $\Delta_1=0$ and the random spin orbit term $\Delta_{4j}=0$. Consequently, the line of fixed points found for class DIII [see Eq.~(\ref{DIIILFP})] is also present within the class AII. But it is unstable in the presence of random chemical potential coupling $\Delta_0$. Additionally, the RG equations for class AII supports a new line of fixed points
\begin{eqnarray}
\Delta^\ast_1+\Delta^\ast_{45}=\frac{3\epsilon}{8}, \; \Delta_4=\Delta_{4j}=0,\label{AIIfp}
\end{eqnarray}
controlled by the random scalar and axial chemical potentials, which controls the quantum phase transitions in the presence of generic time reversal symmetry preserving disorder. This line of fixed points has a universal dynamical scaling exponent $z^\ast=(3-\alpha)/2=1+\epsilon/2$, and within the subspace of scalar and axial chemical potential disorder the diagrammatic perturbation theory will suggest $z^\ast=(3-\alpha)/2$ (for Gaussian white noise disorder $z^\ast=3/2$) to be an exact result. However along this line of fixed points, the anomalous dimension of Dirac mass continuously varies according to
\begin{equation}
\eta_4=\nu^{-1}_M=1-\frac{\epsilon}{4}+\frac{2}{3}\Delta_{45}^{\ast}
\end{equation}
where $0 <\Delta^{\ast}_{45}<3\epsilon/8$. By contrast, the correlation length exponent for the disorder couplings is universal and it is given by $\nu_\Delta=1/\epsilon=1/(1-\alpha)$ (an approximate result of one loop calculation). By setting $Z_{\Psi}=e^{-(3+\eta_\Psi)l}$ we find the corresponding anomalous scaling dimension of the fermion field to be $\eta_\Psi=3\epsilon/8$, which causes the quasiparticle residue of the massless Dirac fermion to vanish along this line of fixed points.

\subsection{Class AIII at d=3} For class AIII we have four particle-hole symmetry preserving disorder couplings $\Delta_5$, $\Delta_{45}$, $\Delta_A$ and $\Delta_{4j}$ respectively corresponding to Dirac mass, axial chemical potential, random Abelian vector potential and random spin orbit potential [see Eq.~(\ref{disorderAIII})]. The RG flow equations are determined by using the scaling dimensions
\begin{eqnarray}
\eta_5&=&\eta_{45}=1-\frac{4}{3}\Delta_5+\frac{4}{3}\Delta_{45}+\frac{8}{3}\Delta_{A}-\frac{8}{3}\Delta_{4,j}, \\
\eta_A&=&\eta_{4,j}=1,\\
\eta_1&=&z(l)=1+\frac{4}{3}\Delta_{45}+\frac{2}{3}\Delta_{5}+\frac{8}{3}\Delta_{A}+\frac{10}{3}\Delta_{4j}. \nonumber \\ 
\end{eqnarray}
The explicit forms of the flow equations are given by
\begin{eqnarray}
&&\frac{d\Delta_5}{dl}=\Delta_5 \left[-\epsilon-\frac{8}{3}\Delta_5 +\frac{8}{3}\Delta_{45}+\frac{16}{3}\Delta_{A}-\frac{16}{3}\Delta_{4j}\right], \nonumber \\  \\
&&\frac{d\Delta_{45}}{dl}=\Delta_{45} \left[-\epsilon-\frac{8}{3}\Delta_5 +\frac{8}{3}\Delta_{45}+\frac{16}{3}\Delta_{A}-\frac{16}{3}\Delta_{4j}\right], \nonumber \\ \\
&&\frac{d\Delta_A}{dl}=-\epsilon \; \Delta_{A},\\
&&\frac{d\Delta_{4j}}{dl}=-\epsilon \; \Delta_{4j}, \\
&&\frac{dm}{dl}=m\left[1-\frac{4}{3}\Delta_5+\frac{4}{3}\Delta_{45}+\frac{8}{3}\Delta_{A}-\frac{8}{3}\Delta_{4j}\right].
\end{eqnarray} 
Clearly the vector potential and spin orbit couplings are irrelevant perturbations and the RG equations for class AIII support a line of fixed points
\begin{equation}
\Delta^\ast_{45}-\Delta^\ast_{5}=\frac{3\epsilon}{8}, \Delta_A=\Delta_{4,j}=0.
\end{equation}
The critical properties of this line of fixed points are identical to the ones found for class DIII. We find $\nu_M=1/\eta_5=2/(3-\alpha)$, $\nu_\Delta=1/\epsilon=1/(1-\alpha)$, $z^\ast=(3-\alpha)/2+ 2 \Delta^\ast_5$, $\eta_\Psi=3\epsilon/8+2\Delta^\ast_5$. Therefore, Gaussian white noise disorder for both symmetry classes DIII and AIII lead to $\nu_M=2/3$, saturating the bound obtained from CCFS theorem~\cite{CCFS}.

\section{Superuniversality in d=4}\label{sec:superd4} We can also carry out a similar RG calculation in $d > 3$. For simplicity we will consider the class AII model at $d=4$, which can be described in terms of a four component massive (TRS preserving Dirac mass) Dirac fermion. By choosing a TRS breaking Dirac mass for four component Dirac fermions we can also describe a four dimensional model of quantum Hall plateau transition belonging to class A. The effective Hamiltonian operator for the clean system in class AII is given by
\begin{equation}
\hat{h}_{4} = v \sum_{j=1}^{4}k_j \Gamma_j + (M-Bk^2)\Gamma_5,
\end{equation}
where we assume $\Psi^\dagger \Gamma_j \Psi$ (with $j=1,2,3,4$) simultaneously breaks inversion and time reversal symmetries, while $\Psi^\dagger \Gamma_5 \Psi$ preserves both symmetries. The inversion symmetry is implemented through $\Psi \to \Gamma_5 \Psi$ and $\mathbf{k} \to -\mathbf{k}$, where $\mathbf{k}$ is the four-dimensional wavevector. The topological invariant for this system follows from $\Pi_4(S^4)=Z$. We can allow the following generic time reversal symmetric disorders
\begin{equation}
H_{d4}=\int d^4x \Psi^\dagger[V_0(\mathbf{x}) \mathbb 1+ V_5(\mathbf{x}) \Gamma_5 + i \sum_{j=1}^{4} V_{5j}(\mathbf{x}) \Gamma_5 \Gamma_j]\Psi.
\end{equation}
The derivation of the beta functions for this model is provided in Appendix~\ref{sec:AppendixB}. The flow equations are given by
\begin{eqnarray}
&&\frac{d\Delta_0}{dl}=\Delta_0 \left[-\epsilon +\frac{5}{2}\Delta_0+\frac{3}{2} \Delta_5+7\Delta_{5j}\right], \\
&&\frac{d\Delta_5}{dl}=\Delta_5 \left[-\epsilon -\frac{3}{2}\Delta_0-\frac{5}{2} \Delta_5+7\Delta_{5j}\right], \\
&&\frac{d\Delta_{5j}}{dl}=\Delta_{5j} \left[-\epsilon +\frac{3}{2}\Delta_0-\frac{3}{2} \Delta_5-3\Delta_{5j}\right]. 
\end{eqnarray}
Apart from the clean massless Dirac fixed point, these equations support a disorder controlled repulsive fixed point (instead of a line of fixed points found for $d=3$)
\begin{equation}
\Delta^\ast_0=\frac{2\epsilon}{5}, \; \Delta_5=\Delta_{5j}=0.\label{FP4d}
\end{equation}
The dynamic scaling exponent at this fixed point is given by 
$z^\ast=1+\epsilon/2$. The Gaussian white noise disorder corresponds to $\epsilon=(d-2)=2$, and leads to $z^\ast=2$. However, the correlation length exponent for disorder $\nu_\Delta=1/\epsilon$ acquires the mean-field value $1/2$ for Gaussian white noise. This indicates $d=4$ to be the upper critical dimension for such disorder driven semimetal-metal transition. Motivated by this fact a Gross-Neveu-Yukawa theory of $Q$ matrix has been proposed in Ref.~\onlinecite{Pixley2}.
The flow equation for the Dirac mass is given by
\begin{equation}
\frac{dm}{dl}=m \; \eta_5=m[1-\frac{3}{4}\Delta_0-\frac{5}{4}\Delta_5+\frac{7}{2}\Delta_{5j}].
\end{equation}
The scaling dimension of Dirac mass at this fixed point becomes $\eta_5=1-3\epsilon/10$. Unless $\epsilon< 10/13$, the disorder scaling dimension $\epsilon >\eta_5$ and we expect the nature of the localization-delocalization transition to be governed by $\nu_\Delta$. Whether the transitions are indeed mean-field in nature (with $z=2$ and $\nu=1/2$) with logarithmic violations of scaling for Gaussian white noise disorder can be checked in future numerical work. However, based on our RG calculations we argue for the stability of superuniversality at weak disorder and its demise at the multicritical point, which leads to a phase diagram similar to the one in Fig.~\ref{phasediagram}. We have also carried out a similar RG calculation for the quantum Hall plateau transition in class A, and found that the multicritical properties are still governed by the fixed point of Eq.~(\ref{FP4d}), even in the presence of additional time reversal symmetry breaking disorder potentials.

\section{Scaling properties of disorder controlled fixed points}\label{sec:scaling}
Next we consider the scaling properties of physical quantities as suggested by the one loop RG equations. For simplicity we mainly discuss class DIII. The scaling behaviors for other two classes can be derived in a similar manner. The solutions to the flow equations for the disorder couplings are given by
\begin{eqnarray}
\Delta_{-}(l)&=&\frac{\Delta^{\ast}_{-}}{1+\left(\frac{\Delta^\ast_- }{\Delta_-(0)}-1\right) e^{ \epsilon l} }, \\
\Delta_{+}(l)&=& \frac{\Delta_+(0)}{\Delta_-(0)} \Delta_-(l),
\end{eqnarray}
where $\Delta_{\pm}=\Delta_{45} \pm \Delta_4$, $\epsilon=1-\alpha$, $\Delta^\ast_-=3\epsilon/4$, and $\Delta_\pm(0)$ are the bare values of the coupling constants. When $\Delta_-(0)>\Delta^\ast_-$), the renormalized disorder couplings diverge beyond a scale $l^\ast_\Delta$. Beyond this scale, the RG equations derived for the ballistic Dirac fermions break down, and the massless Dirac fermion undergoes a quantum phase transition to the diffusive state (DIII-DM in Fig.~\ref{phasediagram}), characterized by a finite density of states at zero energy i. e., $\rho(E=0) \neq 0$. This scale defines the diverging correlation length
\begin{equation}
\xi_\Delta=\Lambda^{-1}e^{l^\ast_\Delta}= \Lambda^{-1} \delta^{-\nu_\Delta},
\end{equation} for disorder with $\delta=(\Delta_-(0)-\Delta^\ast_-)/\Delta^\ast_-$ being the reduced distance from the critical coupling, and $\nu_\Delta=1/\epsilon$. This correlation length can be interpreted as the mean free path of the quasiparticles inside the diffusive phase. The quasiparticle life-time depends on the dynamical exponent $z^\ast=1+\Delta_+(0)$, and
\begin{equation}
\tau \sim \xi^{z^\ast}_{\Delta} \sim \delta^{-\nu_\Delta z^\ast}.
\end{equation}
Notice that the effective exponent $\nu_\Delta z^\ast$ is also nonuniversal. 

After solving the recursion relation for Dirac mass $m$ we obtain the RG invariant
\begin{equation}
\frac{m(l) |\Delta_-(l)|^{\frac{1}{\epsilon}}}{|\Delta_-(l)-\Delta^\ast_-|^{(\frac{1}{2}+\frac{1}{\epsilon})}}=\frac{m(0) |\Delta_-(0)|^{\frac{1}{\epsilon}}}{|\Delta_-(0)-\Delta^\ast_-|^{(\frac{1}{2}+\frac{1}{\epsilon})}}.\label{eqC1}
\end{equation} Therefore, the metal-insulator phase boundary will be determined by
\begin{equation}
\frac{m(0) |\Delta_-(0)|^{\frac{1}{\epsilon}}}{|\Delta_-(0)-\Delta^\ast_-|^{(\frac{1}{2}+\frac{1}{\epsilon})}}=\frac{m(0) |(1+\delta)|^{\frac{1}{\epsilon}}}{(\Delta^\ast_-)^{\frac{1}{\epsilon}}\delta^{(\frac{1}{2}+\frac{1}{\epsilon})}}=C 
\end{equation}
where $C$ is a constant. For $\delta \ll 1$ we can rewrite this in a more elegant form
\begin{equation}
|m(0)| \delta^{-\frac{\nu_\Delta}{\nu_M}}= C (\Delta^\ast_-)^{\nu_\Delta},\label{eqC2}
\end{equation} where $\nu_M=2/(3-\alpha)=2/(2+\epsilon)$.
The scaling behavior of the localization-delocalization transition will be governed by a certain power of the reduced distance from the phase boundary. Since Dirac mass is a more relevant perturbation than the disorder (for any $\alpha>-1/2$), we expect that the scaling phenomena for the localization delocalization transition will be mainly governed by the localization length exponent $\nu_M$. 

Inside the diffusive phase, the thermodynamic and transport properties resemble those of a diffusive Fermi liquid. We emphasize that the class DIII diffusive metal (DIII-DM in Fig.~\ref{phasediagram}) is not an ordinary diffusive metal/Fermi liquid. This state is still a superfluid or superconductor as the total number conservation or global U(1) symmetry remains broken in this phase. Therefore, it is very easy to distinguish a DIII-DM from an ordinary diffusive metal based on the conventional mass or charge transport and Meissner effect. The specific heat and the longitudinal thermal conductivity at low temperatures respectively vary as $C \sim T$ and $\kappa_{xx} \sim T$. We also note that the physical properties of the DIII-DM can alternatively be understood in terms of nonlinear sigma model by following Ref.~\onlinecite{Senthil}. By contrast, the ratio $\kappa_{xx}/(k_BT)$ will vanish exponentially inside the localized phase. When the temperature exceeds $\xi^{-z}_{\Delta}$ and $\xi^{-z}_{M}$, we are in the critical regime of the multicritical line of fixed points, and $C \sim T^{3/z^\ast}$ $\kappa \sim T^{(z^\ast+1)/z^\ast}$. Since $z^\ast$ varies continuously, the power law for $\kappa$ is nonuniversal. However for a model with only random axial chemical potential, our calculations show $z^{-1}=\nu_M=2/(3-\alpha)$. Therefore in the presence of only axial disorder, the thermal conductivity in the critical region behaves universally as
$$\kappa \sim T^x, \: \: x=\frac{5-\alpha}{3-\alpha}.$$ For Gaussian white noise disorder, this exponent becomes $5/3$. By contrast, the thermal conductivity of a Dirac semimetal behaves as $T^2$. The corresponding crossover scaling functions are discussed in Appendix~\ref{sec:AppendixC}.

\begin{figure}[htb]
\begin{center}
\includegraphics[scale=0.65]{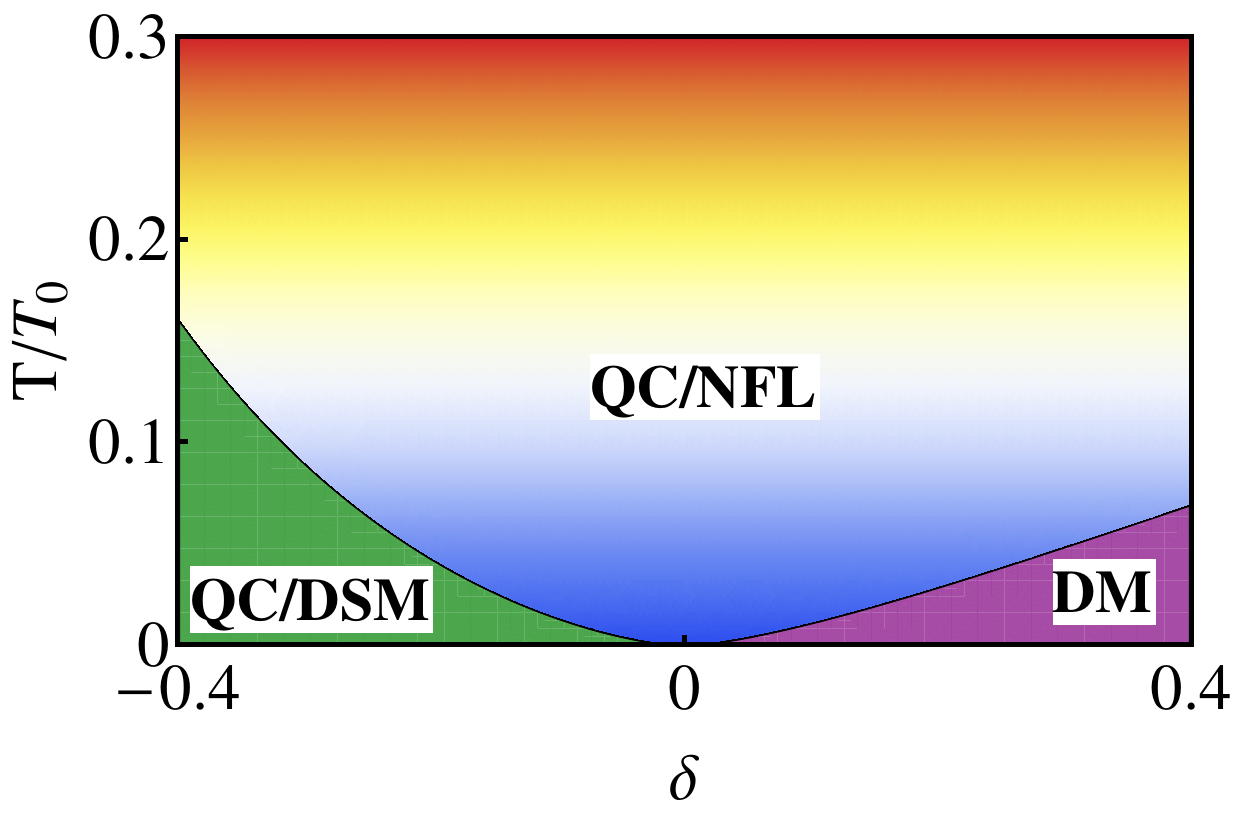}
\caption{Quantum critical fan for the semimetal-metal transition controlled by the multicritical point, in the presence of only random axial chemical potential. The depicted quantum critical fan follows the analytical formula in Eq.~(\ref{fanqc}) obtained for Gaussian white noise distribution. For $\delta<0$ it describes the crossover between the non-Fermi liquid quantum critical region (QC/NFL) and the massless Dirac quantum critical region (QC/DSM). By contrast, for $\delta>0$ we have crossovers between the QC/NFL region and the diffusive metal (DM). If we consider axial chemical potential following Gaussian white noise distribution, the thermal conductivity in QC/NFL, QC/DSM and DM regions respectively follow the power laws $\kappa \sim T^{5/3}$, $\kappa \sim T^2$ and $\kappa \sim T$. A similar fan can be drawn for the direct transition between two localized states occurring at this fixed point.}
\label{fan2}
\end{center}
\end{figure}

Finally we comment on the non-Fermi liquid nature of the multicritical line of fixed points. If we follow the semimetal-metal transition from the weak disorder side the quasiparticle residue of the Dirac fermion vanishes as $\delta^{\eta_\Psi \nu_\Delta}$ (see Ref.~\onlinecite{Pixley2}). Similarly the residue of the diffusons vanishes upon approaching the critical point from the diffusive metal according to $\delta^{\nu_\Delta z}$. The influence of this line of fixed points can be seen over a wide range of temperatures~\cite{Pixley2}, i.e., in the quantum critical fan, as shown in Fig.~\ref{fan2}. This quantum critical fan governs the scaling properties associated with the semimetal-metal transition (if we keep $m=0$ and tune disorder strength). The fan in Fig.~\ref{fan2} is obtained by integrating the RG flow equation for axial disorder ($\Delta_{45}$) following Gaussian white noise distribution and the flow equation for temperature $\frac{dT}{dl}=z(l)T(l)$ with $z(l)$ being the scale dependent dynamic scaling exponent. The details of this method is discussed in Appendix~\ref{sec:AppendixC}. In Fig.~\ref{fan2} we have denoted the reduced distance from the fixed point as $\delta=(\Delta_{45}-\Delta^{\ast}_{45})/\Delta^{\ast}_{45}$. The analytical formula for the quantum critical fan is determined by
\begin{equation}
\frac{T}{T_0} \geq \frac{2}{3\sqrt{3}} \; \frac{|\delta|^{3/2}}{1+\delta},\label{fanqc}
\end{equation}
which gives rise to the anisotropic shape of the fan in Fig.~\ref{fan2}. We also emphasize that the multicritical point denotes the maximum strength of disorder for which a direct transition between two localized states is allowed and the quantum critical fan for this transition is similar to the one in Fig.~\ref{fan}, but with a modified value of $\nu_M z^\ast$. For a generic model of mass and axial disorders, this exponent continuously varies by depending on the strength of mass disorder. Only for pure axial disorder $z^\ast=\nu^{-1}_M=(3-\alpha)/2$, and even at the multicritical point $z^\ast \nu_M$ remains one.

Now we briefly touch upon the scaling properties for AZ symmetry classes AII and AIII in the quantum critical regions. For class AII, there is a universal dynamical exponent $z=(3-\alpha)/2$ along the line of fixed points [see Eq.~(\ref{AIIfp})] controlled by the combination of random scalar and axial potentials. Consequently, the thermal conductivity in QC/NFL, QC/DSM and DM regions (for Gaussian noise distribution) respectively vary as $\kappa \sim T^{5/3}$, $\kappa \sim T^2$ and $\kappa \sim T$. In addition, the temperature variation of the dc charge conductivity in these three regions can be obtained by using $\sigma_c \sim \kappa/T $. The thermal conductivities for class AIII and DIII have identical properties. The charge or spin (when the system describes U(1) spin rotational symmetric paired state) conductivity in class AIII can again be extracted by using $\sigma_{c/s} \sim \kappa/T $.

\section{Conclusion} \label{sec:conclusions} We have considered the effects of quenched disorder on three dimensional, gapped topological states. The direct quantum phase transitions between topologically distinct states for all five symmetry classes are governed by massless Dirac fermion fixed point with emergent continuous chiral symmetry. This is a spectacular example of superuniversality. Based on both perturbative and nonperturbative calculations we have argued for its stability against weak disorder. 

At the perturbative level, we have established a generalized Harris criterion which can be applied for all types of disorder. As the scaling dimension of all fermion bilinears at the massless Dirac fixed point satisfies $\eta <d/2=3/2$, the superuniversality remains robust for sufficiently weak Gaussian white noise disorder. By considering nonperturbative rare events for weak particle-hole symmetric disorder, we have found that the disorder induced Lifshitz tail destroys any sharp spectral gap. The induced low energy density of states  varies as $-\log \rho (E) \sim |E|^{-3}$ and $\rho(E)$ vanishes at zero energy. We have argued that these states are localized with the localization length being determined by the disorder averaged Dirac mass. Therefore in the weak disorder regime, the massless Dirac fermion fixed point controls a direct topological quantum phase transition between two topologically distinct localized phases, with dynamical exponent $z=1$ and the localization length exponent $\nu_M=1$. The renormalization group analysis shows that the superuniversality is absent beyond a critical strength of disorder. For stronger disorder, two localized states are separated by a delocalized diffusive phase. Therefore, the robustness of superuniversality for weak disorder and its eventual demise at a stronger disorder suggests a similar structure for the global phase diagram for all five symmetry classes. Our arguments and calculations are also applicable to any spatial dimension $d > 3$, and suggest the similar structure of the phase diagram and the fate of superuniversality. We have also performed some explicit calculations for topological quantum phase transition within the class AII at $d=4$ for supporting our claims.

Since topological properties are tied to the underlying spatial dimensionality and the symmetry properties of the Dirac Hamiltonian, it is imperative to develop renormalization group analysis at fixed spatial dimensions. We have noted that the disorder with inverse square power-law correlations ($1/r^2$) act as a marginal perturbation for the massless Dirac fixed point, irrespective of the underlying spatial dimensionality. Here, we have introduced disorder distributions with general power-law correlations $1/r^{d-\alpha}$, and developed an $\alpha=\alpha_m-\epsilon$ expansion scheme with $\alpha_m=(d-2)$ for fixed spatial dimensions $d>2$. This allows us to study long range correlated disorder and by taking the formal limit $\epsilon \to (d-2)$ we have also made new predictions for Gaussian white noise disorder distribution. We have also elaborated on the subtleties involved with taking such limits for the leading order analysis.

Our work should stimulate nonperturbative analytical and numerical analysis of the topological quantum phase transitions for different symmetry classes for $d \geq 3$. On the experimental side, there are some data for topological insulator to trivial insulator transition for spin orbit coupled materials. However due to the presence of a metallic bulk it is hard to obtain reliable information regarding the massless Dirac fermion nature of the critical point. We anticipate that some of our predictions regarding class DIII can be tested for $^3$He B phase on porous vycor and aerogel. The pseudoscalar pairing of four component charged Dirac fermions~\cite{FuBerg} is an experimentally pertinent example of TSC in the class DIII. Upon projection onto the low energy quasiparticles in the vicinity of the Fermi surface, the effective Hamiltonian for pseudoscalar pairing maps onto the Hamiltonian of $^3$He B-phase~\cite{Silaev}. Following the suggestion of Ref.~\onlinecite{FuBerg}, there has been experimental interest in the superconducting $\mathrm{Cu}_x\mathrm{Bi}_2\mathrm{Se}_3$ and $\mathrm{Sn}_{1-x}\mathrm{In}_x\mathrm{Te}$~(Ref.~\onlinecite{ObWray1, ObKriener, ObWray2,ObSasaki,PCSasaki, PCKirzhner, PCChen, ObLevy, Novak, Tranquada}). The actual topological nature of the paired state in this material is still under debate. However recent NMR experiments have provided evidence for triplet pairing, with the $d$ vector being locked in the $ab$ plane. Such a state can still be topological~\cite{Fu}, but it does not correspond to the pseudoscalar pairing of Dirac fermions. Since these materials are quite dirty, disorder can have substantial effects toward determining the actual nature of the paired state.\\

\emph{Note Added}: Recent numerical calculations of single particle propagator for Weyl Hamiltonian in the presence of scalar potential disorder~\cite{PixleyNandkishore} has found $Z(k) \sim k^{0.4}$ and $G(k) \sim k^{-1.13}$, in good agreement with our one loop results $Z(k) \sim k^{0.375}$ and $G(k) \sim k^{-1.125}$ obtained from the expansion scheme (I).\\

\acknowledgments

P. G. was supported by JQI-NSF-PFC and LPS-MPO-CMTC. S. C. was supported by funds from David S. Saxon Presidential Term Chair at UCLA. We thank S. Das Sarma, B. Roy and J. Pixley for discussions.

\onecolumngrid

\appendix

\begin{figure*}[htb]
\begin{center}
\includegraphics[scale=0.6]{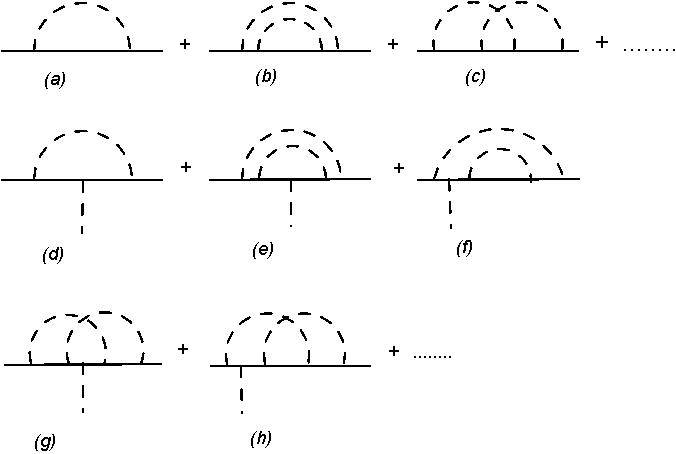}
\caption{Feynmann diagrams needed for calculating anomalous scaling dimension of a fermion bilinear. The vertex diagrams can be obtained by differentiating the self-energy with respect to the appropriate source term as a consequence of Ward identity.}
\label{fa}
\end{center}
\end{figure*}

\section{Scaling dimensions for axial chemical potential} \label{sec:AppendixA} We elaborate upon the algebraic reason why the anomalous scaling dimensions of the operators $\Psi^\dagger \mathbb 1 \Psi$, $\Psi^\dagger \Gamma_4 \Psi$, $\Psi^\dagger \Gamma_5 \Psi$ and $\Psi^\dagger \Gamma_{45} \Psi$ ($\Gamma_{45}=i \Gamma_4 \Gamma_5$) in the presence of only axial chemical potential disorder are equal. The perturbative corrections to the operators have to be calculated by evaluating Feynmann diagrams similar to those in panels (d), (e), (f), (g) and (h) in Fig.~\ref{fa}. Here the bare three point vertex is created by an appropriate source term that couples to one of the operators $\Psi^\dagger \hat{M} \Psi$, where $\hat{M}=\mathbb 1, \Gamma_4, \Gamma_5, \Gamma_{45}$, while dressing of the vertex will be carried out with the help of axial disorder vertices.

For the one loop diagram of Fig.~(6d), the integrand will involve a matrix product $\Gamma_{45}\Gamma_j \hat{M} \Gamma_l \Gamma_{45}$. The free indices of the Dirac matrices will remain contracted with momentum components inside the loop integral. Due to the properties $[\Gamma_j,\Gamma_{45}]=0$, $\{ \Gamma_j, \Gamma_4 \}=0$, $\{ \Gamma_j, \Gamma_5 \}=0$, $\{ \Gamma_4, \Gamma_{45} \}=0$, and $\{ \Gamma_5, \Gamma_{45} \}=0$, $\Gamma_{45}\Gamma_j \hat{M} \Gamma_l \Gamma_{45}=\Gamma_j\Gamma_l \hat{M}$ for any of the above four operators. Therefore $\Psi^\dagger \hat{M} \Psi$ receives exactly the same perturbative correction at one loop level. Therefore, $\eta_1=\eta_4=\eta_5=\eta_{45}$ at one loop order.

For the two loop diagram of Fig.~(6e), the integrand involves the matrix product $\Gamma_{45}\Gamma_j\Gamma_{45}\Gamma_l\hat{M}\Gamma_{n}\Gamma_{45}\Gamma_{k}\Gamma_{45}$. From the commutation properties of the participating matrices we find $\Gamma_{45}\Gamma_j\Gamma_{45}\Gamma_l\hat{M}\Gamma_{n}\Gamma_{45}\Gamma_{k}\Gamma_{45}=\Gamma_j\Gamma_l\Gamma_n\Gamma_k \hat{M}$. Therefore the two loop Feynmann diagram like Fig.~(6e) leads to the identical perturbative corrections for $\hat{M}=\mathbb 1, \Gamma_4, \Gamma_5, \Gamma_{45}$. This holds for other two loop diagrams of Fig.~(6f), Fig.~(6g) and Fig.~(6h) as well. Also proceeding in the similar manner with the higher loop diagrams, we can show that at any order of perturbation theory the corrections to $\Psi^\dagger \hat{M} \Psi$ with $\hat{M}=\mathbb 1, \Gamma_4, \Gamma_5, \Gamma_{45}$ due to axial disorder are equal. Therefore, for axial disorder model, $\eta_1=\eta_4=\eta_5=\eta_{45}$ holds at any order of perturbation theory. Consequently, $z=\nu^{-1}_M$ at the axial disorder controlled multicritical point ($z=3/2$ for Gaussian white noise distribution).

We can put this line of reasoning on a stronger footing by performing symmetry analysis of the replica field theory at zero energy. Since disorder only causes elastic scattering we are allowed to consider different values of energy separately. In the presence of only axial disorder the replicated effective action for massless Dirac fermions at zero energy is given by  
\begin{equation}
S_{ax}=\int d^3x \Psi^\dagger_a [-i  v \; \sum_{j=1}^{3} \; \Gamma_j  \partial_j + \sqrt{\Delta_{45}} \; V_{45}(\mathbf{x}) \Gamma_{45}]\Psi_a + \frac{1}{2} \; \int d^3x \; V_{45}(-\nabla^2)^{\alpha/2}V_{45},
\end{equation}
where $a$ is the replica index. We have used a disorder propagator which in the momentum space behaves as $|\mathbf{q}|^\alpha$. Alternatively we can use $\int d^3x \; d^3y V_{45}(\mathbf{x}) V_{45}(\mathbf{y}) |\mathbf{x}-\mathbf{y}|^{-(d+\alpha)}$. The effective action is invariant under the following transformations
\begin{eqnarray}
&& \Psi \to e^{i \phi_0 \mathbb 1} \Psi, \; \Psi^\dagger \to \Psi^\dagger e^{-i \phi_0 \mathbb 1}; \: \Psi \to e^{i \phi_{45} \Gamma_{45}} \Psi, \; \Psi^\dagger \to \Psi^\dagger e^{-i \phi_{45} \Gamma_{45}}; \nonumber \\
&& \Psi \to e^{i \phi_4 \Gamma_4} \Psi, \; \Psi^\dagger \to \Psi^\dagger e^{i \phi_4 \Gamma_4}; \: \Psi \to e^{i \phi_{5} \Gamma_{5}} \Psi, \; \Psi^\dagger \to \Psi^\dagger e^{i \phi_{5} \Gamma_{5}},
\end{eqnarray}
which give rise to $U(2)$ chiral symmetry for a single replica index. By combining with the replica index this leads to $U(2n)$ chiral symmetry. When the disorder strength $\Delta_{45} \leq \Delta^\ast_{45}$, $U(2n)$ symmetry remains unbroken. Inside the diffusive phase a constant scattering rate or finite density of states breaks this symmetry down to $U(n) \times U(n)$ corresponding to the transformations
\begin{eqnarray}
&& \Psi \to e^{i \phi_0 \mathbb 1} \Psi, \; \Psi^\dagger \to \Psi^\dagger e^{-i \phi_0 \mathbb 1}; \: \Psi \to e^{i \phi_{45} \Gamma_{45}} \Psi, \; \Psi^\dagger \to \Psi^\dagger e^{-i \phi_{45} \Gamma_{45}}. 
\end{eqnarray}
The equality of the scaling dimensions for $\Psi^\dagger \mathbb 1 \Psi$, $\Psi^\dagger \Gamma_4 \Psi$, $\Psi^\dagger \Gamma_5 \Psi$ and $\Psi^\dagger \Gamma_{45} \Psi$ at the axial disorder controlled fixed point is a consequence of the unbroken $U(2n)$ chiral symmetry.

\section{Derivation of renormalization group flow equations}\label{sec:AppendixB}
We will first consider in Subsec.~\ref{subsec1} the derivation of beta functions for class AII in $d=3$ for $\alpha>0$ by employing scheme (I). The beta functions for class DIII can be obtained by setting coupling constants of random scalar potential ($\Delta_0$) and random spin orbit potentials ($\Delta_{4j}$) to be zero. The flow equations for class AIII in $d=3$ and class AII in $d=4$ are respectively derived in the Subsec.~\ref{subsec2} and Subsec.~\ref{subsec3}. Since, the one loop ladder and crossing diagrams of 
Fig.~3(c) and 3(d) are UV convergent for $\alpha>0$ and $d>2$, they do not contribute to the beta functions. Consequently, we can use diagrams of Fig.~3(a) and 3(b) or equivalently one loop diagrams of Fig.~\ref{fa} for obtaining the one loop beta functions, when $\alpha>0$ and $d>2$.
 
\subsection{Class AII and Class DIII in three dimensions} \label{subsec1}After ignoring the velocity anisotropies and higher gradient terms, the replicated effective action is given by
\begin{eqnarray}
S&=&\int d^3x d\tau \Psi^\dagger_a \bigg[\partial_\tau -i v \; \sum_{j=1}^{3} \; \Gamma_j  \partial_j + m \Gamma_4 + \sqrt{\Delta_0} \; V_0(\mathbf{x}) \mathbb{1} + \sqrt{\Delta_4} \; V_4(\mathbf{x}) \Gamma_4 \nonumber \\ && + \sqrt{\Delta_{45}} \; V_{45}(\mathbf{x})\Gamma_{45} + \sqrt{\Delta_{4j}} \; \sum_{j=1}^{3} \; V_{4j}(\mathbf{x}) \Gamma_{4j} \bigg]\Psi_a 
+ \frac{1}{2}\int d^3x \bigg[V_0(-\nabla^2)^{\alpha/2}V_0 \nonumber \\ && +V_4(-\nabla^2)^{\alpha/2}V_4 + V_{45}(-\nabla^2)^{\alpha/2}V_{45} +\sum_{j=1}^{3}V_{4j}(-\nabla^2)^{\alpha/2}V_{4j}\bigg],
\end{eqnarray}
 with $\alpha=(d-2)-\epsilon=1-\epsilon$.  

The bare propagator of the fermion field is given by
\begin{equation}
G_0(i\omega,\mathbf{k})=-\frac{i\omega -v\Gamma_j k_j -m}{\omega^2+v^2k^2+m^2}.
\end{equation}
The disorder propagator is $D_\mu(\mathbf{q})= |\mathbf{q}|^{1-\epsilon}$, and for the perturbative calculations of UV divergent terms we simply need the integral $\int d^3k/(k^{3-\epsilon}) \sim \int dk/k \sim l= -\log \Lambda$. Now onward we will use dimensionless variables $m \to m/(v \Lambda)$ and $\Delta_a \to \Delta_a \Lambda^\epsilon /(2\pi^2 v^2)$. After some simple matrix algebra, the UV divergent part of the fermion self-energy at one loop level becomes
\begin{eqnarray}
\Sigma(i\omega, \mathbf{k})=-i\omega(\Delta_0+\Delta_4+\Delta_{45}+3\Delta_{4j})l+\frac{1}{3} v\Gamma_j k_j (\Delta_0+\Delta_{45}+\Delta_{4j}-\Delta_4)l \nonumber \\ +m(\Delta_0+\Delta_4-\Delta_{45}-3\Delta_{4j})l.
\end{eqnarray}
The renormalized propagator will be given by $G^{-1}=G^{-1}_0-\Sigma$. Therefore, $\partial_\tau$ will have a coefficient $[1+(\Delta_0+\Delta_4+\Delta_{45}+3\Delta_{4j})l]$, while the coefficient of kinetic energy term $v \Gamma_j k_j$ is given by $\left[1-\frac{1}{3}(\Delta_0+\Delta_{45}+\Delta_{4j}-\Delta_4)l \right]$. Similarly the coefficient of Dirac mass becomes $[1-(\Delta_0+\Delta_4-\Delta_{45}-3\Delta_{4j})l]$. After the rescaling $x \to x e^l$, $\tau \to \tau e^{zl}$ and $\Psi \to \Psi Z^{1/2}_{\Psi}$ we find the scaling dimensions reported in Eq.~(\ref{AIIbeta1}), Eq.~(\ref{AIIbeta2}), and Eq.~(\ref{AIIbeta5}). After the perturbative vertex corrections, the Yukawa couplings for potential, mass, axial and spin orbit disorders respectively modify to 
\begin{eqnarray}
&& \sqrt{\Delta_0}\int d^3x d\tau \Psi^\dagger_a \mathbb{1} \Psi_a V_0(\mathbf{x}) \left[1+\Delta_0 l+\Delta_4 l+\Delta_{45} l+3 \Delta_{4j} l \right], \\
&& \sqrt{\Delta_4}\int d^3x d\tau \Psi^\dagger_a \Gamma_4 \Psi_a V_4(\mathbf{x}) \left[1-\Delta_0 l-\Delta_4 l+\Delta_{45} l+3 \Delta_{4j} l \right],\\
&& \sqrt{\Delta_{45}}\int d^3x d\tau \Psi^\dagger_a \Gamma_{45} \Psi_a V_{45}(\mathbf{x}) \left[1+\Delta_0 l-\Delta_4 l+\Delta_{45} l-3 \Delta_{4j} l \right],\\
&& \sqrt{\Delta_{4j}}\int d^3x d\tau \Psi^\dagger_a \Gamma_{4j} \Psi_a V_{4j}(\mathbf{x}) \left[1+\frac{1}{3}\Delta_0 l-\frac{1}{3}\Delta_4 l-\frac{1}{3}\Delta_{45} l-\frac{1}{3} \Delta_{4j} l \right].
\end{eqnarray} 
Under the rescaling $V_{\mu}=Z^{1/2}_{V,\mu} V_{\mu}$ with $Z_{V,\mu}=e^{-(d-\alpha) l}=e^{-(3-\alpha) l}$, we find the other two scaling dimensions in Eq.~(\ref{AIIbeta3}) and Eq.~(\ref{AIIbeta4}). After setting $\Delta_0=\Delta_{4j}=0$ we find the scaling dimensions for class DIII. 

\subsection{Class AIII in three dimensions}\label{subsec2}
For class AIII the replicated effective action is given by
\begin{eqnarray}
S&=&\int d^3x d\tau \Psi^\dagger_a \bigg[\partial_\tau -i v \; \sum_{j=1}^{3} \; \Gamma_j  \partial_j + m \Gamma_5 + \sqrt{\Delta_5} \; V_5(\mathbf{x}) \Gamma_5 + \sqrt{\Delta_{45}} \; V_{45}(\mathbf{x})\Gamma_{45} \nonumber \\ && + \sqrt{\Delta_{A}} \; \sum_{j=1}^{3} \; V_{j}(\mathbf{x}) \Gamma_{j}+ \sqrt{\Delta_{4j}} \; \sum_{j=1}^{3} \; V_{4j}(\mathbf{x}) \Gamma_{4j} \bigg]\Psi_a 
+ \frac{1}{2}\int d^3x \bigg[V_5(-\nabla^2)^{\alpha/2}V_5 \nonumber \\ && +V_{45}(-\nabla^2)^{\alpha/2}V_{45}  +\sum_{j=1}^{3} V_{j}(-\nabla^2)^{\alpha/2}V_{j} +\sum_{j=1}^{3}V_{4j}(-\nabla^2)^{\alpha/2}V_{4j}\bigg].
\end{eqnarray}
the ultraviolet divergent part of the fermion self-energy at one loop level becomes
\begin{eqnarray}
\Sigma(i\omega, \mathbf{k})=-i\omega(\Delta_5+\Delta_{45}+3\Delta_{A}+3\Delta_{4j})l+\frac{1}{3} v\Gamma_j k_j (-\Delta_5+\Delta_{45}+\Delta_{4j}-\Delta_A)l \nonumber \\ +m(\Delta_5-\Delta_{45}-3\Delta_A+3\Delta_{4j})l.
\end{eqnarray}
After accounting for the perturbative corrections, the Yukawa vertices become 
\begin{eqnarray}
&& \sqrt{\Delta_5}\int d^3x d\tau \Psi^\dagger_a \Gamma_5 \Psi_a V_5(\mathbf{x}) \left[1-\Delta_5 l+\Delta_4 l+\Delta_{45} l+3 \Delta_{A} l -3\Delta_{4j}l\right], \\
&& \sqrt{\Delta_{45}}\int d^3x d\tau \Psi^\dagger_a \Gamma_{45} \Psi_a V_{45}(\mathbf{x}) \left[1-\Delta_5 l+\Delta_4 l+\Delta_{45} l+3 \Delta_{A} l -3\Delta_{4j}l\right],\\
&& \sqrt{\Delta_{A}}\int d^3x d\tau \Psi^\dagger_a \Gamma_{j} \Psi_a V_{j}(\mathbf{x}) \left[1+\frac{1}{3}\Delta_5 l-\frac{1}{3}\Delta_{45} l+\frac{1}{3}\Delta_{A} l-\frac{1}{3} \Delta_{4j} l \right],\\
&& \sqrt{\Delta_{4j}}\int d^3x d\tau \Psi^\dagger_a \Gamma_{4j} \Psi_a V_{4j}(\mathbf{x}) \left[1+\frac{1}{3}\Delta_5 l-\frac{1}{3}\Delta_{45} l+\frac{1}{3}\Delta_{A} l-\frac{1}{3} \Delta_{4j} l \right].
\end{eqnarray}
Now following the rescaling procedure described in the previous subsection we obtain the required scaling dimensions and the RG flow equations for class AIII.

\subsection{Class AII in four dimensions}\label{subsec3}
The replicated effective action is given by
\begin{eqnarray}
S&=&\int d^4x d\tau \Psi^\dagger_a \bigg[\partial_\tau -i v \; \sum_{j=1}^{4} \; \Gamma_j  \partial_j + m \Gamma_5 + \sqrt{\Delta_0} \; V_0(\mathbf{x}) \mathbb{1} + \sqrt{\Delta_5} \; V_5(\mathbf{x}) \Gamma_5 \nonumber \\ &&  + \sqrt{\Delta_{5j}} \; \sum_{j=1}^{4} \; V_{5j}(\mathbf{x}) \Gamma_{5j} \bigg]\Psi_a 
+ \frac{1}{2}\int d^3x \bigg[V_0(-\nabla^2)^{\alpha/2}V_0  +V_5(-\nabla^2)^{\alpha/2}V_5 \nonumber \\ &&+\sum_{j=1}^{4}V_{5j}(-\nabla^2)^{\alpha/2}V_{5j}\bigg],
\end{eqnarray}
where $\alpha=(d-2)-\epsilon=2-\epsilon$. The Gaussian white noise distribution corresponds to $\alpha=0$ or $\epsilon=2$. The disorder propagator is $|\mathbf{q}|^{2-\epsilon}$ and we need the integral $\int d^4k/(k^{4-\epsilon}) \sim \int dk/k \sim l= -\log \Lambda$. The dimensionless disorder couplings will be denoted as $\Delta_a \to \Delta_a \Lambda^\epsilon/(8\pi^2 v^2)$. The ultraviolet divergent part of the selfenergy at one loop level is given by
\begin{eqnarray}
\Sigma(i\omega, \mathbf{k})=-i\omega(\Delta_0+\Delta_5+4\Delta_{5j})l+\frac{1}{4} v\Gamma_j k_j (-\Delta_5+\Delta_{0}-2\Delta_{5j})l \nonumber \\ +m(\Delta_0+\Delta_{5}-4\Delta_{5j})l.
\end{eqnarray}
After accounting for the perturbative corrections the Yukawa vertices become
\begin{eqnarray}
&& \sqrt{\Delta_0}\int d^4x d\tau \Psi^\dagger_a \mathbb{1} \Psi_a V_0(\mathbf{x}) \left[1+\Delta_0 l+\Delta_5 l+4\Delta_{5j} l \right], \\
&& \sqrt{\Delta_5}\int d^4x d\tau \Psi^\dagger_a \Gamma_5 \Psi_a V_5(\mathbf{x}) \left[1-\Delta_0 l-\Delta_5 l+4\Delta_{5j} l \right], \\
&& \sqrt{\Delta_{5j}}\int d^4x d\tau \Psi^\dagger_a \Gamma_{5j} \Psi_a V_0(\mathbf{x}) \left[1+\frac{1}{2}\Delta_0 l-\frac{1}{2}\Delta_5 l-\Delta_{5j} l \right].
\end{eqnarray}
After the recsaling $x \to x e^l$, $\tau \to \tau e^{zl}$, $\Psi \to Z^{1/2}_\Psi \Psi$, $V_{\mu}=Z^{1/2}_{V,\mu} V_{\mu}$ with $Z_{V,\mu}=e^{-(d-\alpha) l}=e^{-(4-\alpha) l}$ we obtain the required scaling dimensions and the RG flow equations for class AII in four dimensions.

\section{Solution of DIII beta functions}\label{sec:AppendixC}

The constant $C$ can be determined by solving the crossover length scale $l^\ast_M$ such that $m(l) \sim 1$. After combining the results from Eq.~(\ref{eqC1}) and Eq.~(\ref{eqC2}) the explicit $l$ dependence of $m$ is found to be
\begin{equation}
\frac{m(l)}{m(0)}=\sqrt{\frac{\Delta^\ast_-}{\Delta_-(0)}} \; \frac{e^{(1+\epsilon/2)l}}{\sqrt{1+\left(\frac{\Delta^\ast_- }{\Delta_-(0)}-1\right) e^{ \epsilon l}}}.
\end{equation}
When $\Delta_-(0)<\Delta^\ast_-$ and $\left(\frac{\Delta^\ast_- }{\Delta_-(0)}-1\right) e^{ \epsilon l_M}>>1$, the $m(l) \sim 1$ for 
$$e^{l_M} \sim \delta^{1/2}/|m(0)|$$, which gives the disorder dependence of the localization length in the weak disorder limit. In the opposite limit $\left(\frac{\Delta^\ast_- }{\Delta_-(0)}-1\right) e^{ \epsilon l_M}<<1$ we have $\xi_M \sim |m(0)|^{-\nu_M}$, with $\nu^{-1}_{M}=1+\epsilon/2$.
For $d=3$ and $\epsilon=1$, $e^{l_M}$ can be obtained from the roots of a cubic equation. When $\Delta_-(0)<\Delta^\ast_-$ the cubic equation can be written in the form
\begin{equation}
e^{3l^\ast_M}-\frac{\delta}{m^2(0)}e^{l^\ast_M} -\frac{1-\delta}{m^2(0)}=0.
\end{equation} By contrast, for $\Delta^\ast_- <\Delta_-(0)$ the cubic equation becomes
\begin{equation}
e^{3l^\ast_M}+\frac{\delta}{m^2(0)}e^{l^\ast_M} -\frac{1+\delta}{m^2(0)}=0.
\end{equation}
For $\Delta^\ast_- <\Delta_-(0)$ there is only one real root
\begin{equation}
e^{l^\ast_M}=\frac{3 \xi_\Delta \Lambda}{x} \sinh \left(\frac{1}{3} \sinh^{-1} x\right)
\end{equation}
where $$x=\frac{3\sqrt{3}}{2} m(0) \delta^{-3/2},$$ is the dimensionless scaling variable. From this quantity, we can estimate the constant $C= 4\sqrt{3}/9$.  For $\Delta_-(0)<\Delta^\ast_-$ and $x<1$, 
\begin{equation}
e^{l^\ast_M}=\frac{3 \xi_\Delta \Lambda}{x} \cos \left(\frac{1}{3} \cos^{-1} x\right).
\end{equation} For $\Delta_-(0)<\Delta^\ast_-$ and $x>1$, 
\begin{equation}
e^{l^\ast_M}=\frac{3 \xi_\Delta \Lambda}{x} \cosh \left(\frac{1}{3} \cosh^{-1} x\right).
\end{equation} This leads to the following asymptotic behavior for the localization length in the vicinity of the insulator-insulator transition
\begin{equation}
\xi_M \sim \Lambda^{-1} \delta^{1/2} |m(0)|^{-1}
\end{equation} governed by the massless Dirac fermion. Actually the arguments of inverse functions are $x=\frac{3\sqrt{3}}{2} m(0) (1 \pm \delta)\delta^{-3/2}$, depending on $\Delta_-^\ast<\Delta_-(0)$ or $\Delta_-^\ast>\Delta_-(0)$.

Now we analyze the flow equation for the dimensionless variable $E/(v \Lambda)$ or $T/(v \lambda)$ where $E$ and $T$ respectively denote the energy and temperature. We find
\begin{equation}
E(l)=E(0) \left[\frac{\Delta^\ast_-}{\Delta_-(0)}\right]^{y}\frac{\exp \left[l \left \{1+y \right \}\right]}{\left[1+\left(\frac{\Delta^\ast_- }{\Delta_-(0)}-1\right) e^{ \epsilon l} \right]^{y}},
\end{equation} where $y=\frac{3}{8}\left(\frac{\Delta^\ast_- }{\Delta_-(0)}+\frac{1}{3}\right)$. For generic values of the coupling constants the cross-over length $l^\ast_E$ has to be found numerically. However, for the special values (i) $\Delta_+(0)/\Delta_-(0)=1$ corresponding to pure axial disorder ($\Delta_M=0$) and (ii) $\Delta_+(0)/\Delta_-(0)=7/3$ corresponding to a trajectory $\Delta_{4}=(2/5)\Delta_{45}$, we can calculate $l^\ast_E$ analytically.

For $\Delta_+(0)/\Delta_-(0)=1$, the $z^\ast=3/2$, and $l^\ast_E$ becomes identical to $l^\ast_M$ determined above, after replacing $m(0)$ by $E(0)$. For this case and insulator-insulator transition, $|E(0)| > |m(0)|$ is the required condition for observing critical behavior with dynamical exponent $z=3/2$. 
Let us now focus on the case $\Delta_{4}=(2/5)\Delta_{45}$. The scale $l^\ast_E$ is obtained by solving the quadratic equation
\begin{equation}
e^{2 l^\ast_E}- \frac{\Delta^\ast_- - \Delta_-(0)}{\Delta^\ast_- \; |E(0)|}e^{l^\ast_E}-\frac{1}{|E(0)|}=0.
\end{equation} When $\Delta_-(0)=\Delta^\ast_-$, $\Delta_+(0)=1$ and $z^\ast=2$. Therefore, at the critical strength of disorder
\begin{equation}
e^{l^\ast_E}=|E(0)|^{-1/2}.
\end{equation} For $\Delta_-(0)<\Delta^\ast_-$,
\begin{equation}
e^{l^\ast_E}=\frac{\xi_\Delta \Lambda}{|x|}\left[1+\sqrt{1+2|x|}\right],
\end{equation} where $x=2E(0)\delta^{-2}$.  When $x>>1$, we have $z^\ast=2$ critical behavior, which finally gives away to the $z^\ast=1$ scaling behavior of the massless Dirac fermion in the region $x<<1$. For $\Delta_-(0)>\Delta^\ast_-$,
\begin{equation}
e^{l^\ast_E}=\frac{\xi_\Delta \Lambda}{|x|}\left[\sqrt{1+2|x|}-1 \right].
\end{equation} Again this the cross-over scale displays $z^\ast=2$ critical behavior for $x>>1$. When $x<<1$, we find the characteristic behavior of the diffusive metal $e^{l^\ast_E} \sim \xi_\Delta \Lambda$. As before for large $\delta$ there will be the factors $(1\pm \delta)$ in the argument under square root. The cross-over scale for temperature is identical to the one obtained for energy. The explicit form of the scaling function for the density of states is obtained by taking derivative of $e^{-3l^\ast_E}$ with respect to $E$. Subsequently, any thermodynamic quantity can be easily obtained from the singular part of the free energy density $$f \sim \int dE \; E \; \rho(E) \; n_F(E/T), $$ where $n_F(E/T)$ is the Fermi distribution function. For Dirac semimetal-metal transition studied in Ref.~\onlinecite{Pixley2}, we have found very good agreement between such analytically predicted scaling function and the numerically determined one. The scaling function for $\kappa$ can be obtained by taking $\kappa(T)=T e^{-(d-2)l^\ast_T}$, and $l^\ast_T$ is obtained by replacing $E$ with $T$ in the formulas for $l^\ast_E$.

\twocolumngrid

\end{document}